\documentclass[showpacs,preprintnumbers,prd,amssymb]{revtex4}
\begin{document}
\preprint{IFT.P-94/2002}
\title{Metric-affine approach to teleparallel gravity}
\author{Yu.N. Obukhov}
\altaffiliation[Permanent address: ]{Department of 
Theoretical Physics, Moscow State University, 117234 Moscow, Russia}
\author{J.G. Pereira}
\affiliation{Instituto de F\'{\i}sica Te\'orica,
Universidade Estadual Paulista\\
Rua Pamplona 145,
01405-900 S\~ao Paulo SP, Brazil}

\begin{abstract}
The teleparallel gravity theory, treated physically as a gauge theory of 
translations, naturally represents a particular case of the most general 
gauge-theoretic model based on the general affine group of spacetime. On the
other hand, geometrically, the Weitzenb\"ock spacetime of distant
parallelism is a particular case of the general metric-affine spacetime
manifold. These physical and geometrical facts offer a new approach to the
teleparallelism. We present a systematic treatment of the teleparallel
gravity within the framework of the metric-affine theory. The symmetries,
conservation laws and the field equations are consistently derived, and the
physical consequences are discussed in detail. We demonstrate that the
so-called teleparallel GR-equivalent model has a number of attractive
features which distinguishes it among the general teleparallel theories, 
although it has a consistency problem when dealing with spinning 
matter sources.  
\end{abstract}


\maketitle

\section{Introduction}

Theories of gravity based on the geometry of distant parallelism 
\cite{HN,HS,jose1,jose2,jose3,wojtek,nester} are commonly considered as the
closest alternative to the general relativity (GR) theory. Teleparallel
gravity models possess a  number of attractive features both from the
geometrical and physical viewpoints. Teleparallelism naturally arises within
the framework of the gauge theory of the spacetime translation group.
Translations are closely related to the group of general coordinate
transformations which underlies GR. Accordingly, the energy-momentum current
represents the matter source in the field equations of the teleparallel
gravity, like in GR.

Geometrically, the teleparallel models are described by the 
Weitzenb\"ock spacetime. The latter is characterized by the
trivial curvature and non-zero torsion. The tetrad (or a coframe) 
field is the basic dynamical variable which can be treated as
the gauge potential corresponding to the group of local 
translations. Then the torsion is naturally interpreted as
the corresponding gauge field strength. As a result, a gravitational
teleparallel Lagrangian is straightforwardly constructed, in a 
Yang-Mills manner, from the quadratic torsion invariants.

The teleparallel gravity
belongs naturally to the class of the metric-affine gravitational
theories (MAG). Quite generally \cite{PR}, MAG can be understood within 
the framework of the gauge approach for the affine group, a semidirect 
product of the general linear group and the translation group. 
The corresponding gauge field potentials are the linear 
connection and the coframe, whereas the curvature and the torsion
turn out to be the gauge gravitational field strengths. Besides,
the metric represents one more fundamental field with the
nonmetricity as the corresponding field strength. 

{}From the gauge-theoretic point of view, teleparallel gravity is
distinguished among the other MAG models by reducing the affine 
symmetry group to the translation subgroup. In geometric language,
teleparallelism arises from the general metric-affine spacetime
after we impose two constraints by putting curvature and
nonmetricity equal zero. In this sense, teleparallelism can
be treated, both physically and geometrically, as a particular 
case of the general MAG theory. 

In the present paper we will study teleparallel gravity within the MAG 
approach. A similar analysis was previously developed in \cite{wojtek} 
for the Poincar\'e gauge approach.
We will consider the general class of teleparallel models which are 
characterized by three dimensionless coupling constants. After
recalling some basic facts about MAG in Sec.~\ref{mag}, we derive the 
field equations of the general teleparallel theory in Sec.~\ref{eqs}.
Then in Sec.~\ref{ein} we show that the teleparallel gravity model 
can be reduced to an effective Einstein theory with some modified 
energy-momentum current as a source. When the coupling constants  
have specific values (\ref{aGR}), the exact equivalence between general 
relativity and teleparallel gravity is established for spinless 
sources. However, in Sec.~\ref{lim} we demonstrate certain 
consistency problems for the sources with spin. Among the general 
teleparallel models, there is a class of theories with the proper 
conformal properties. We discuss the definition of conformal symmetry
in teleparallelism in Sec.~\ref{co}. Finally, we analyze the 
spherically symmetric solutions for the general teleparallel gravity 
coupled to the electromagnetic field. We pay special
attention to the behavior of the (Riemannian) curvature and torsion
invariants. This aspect was not studied in detail before. The
formulation of the spherically symmetric problem and its preliminary 
analysis is given in Secs.~\ref{ansatz}-\ref{an}. A number of new 
solutions is obtained, both charged and uncharged. In Sec.~\ref{flat} 
we present the class of conformally flat solutions that generalizes 
the solutions of Bertotti and Robinson. The corresponding geometry 
appears to be regular from both the Riemannian curvature and the 
teleparallel torsion points of view. The complete set of uncharged 
solutions is presented in Sec.~\ref{un}. This is done for the first 
time for all possible choices of the three coupling constants of 
teleparallel gravity. We analyze in Sec.~\ref{values} the corresponding 
spacetime geometries, and find that the generic solutions  are no 
black holes. In Sec.~\ref{ch} we show that only in the above mentioned 
general relativity limit, i.e.\ for the specific values of the coupling 
constants (\ref{aGR}), we recover the  Schwarzschild and the 
Reissner-Nordstr\"om black hole configurations. This fact can be considered
as an argument favoring the choice of the rigid structure of the
teleparallel Lagrangian with the fixed values of the three coupling
constants. The analysis of the torsion invariants reveals, however, that
even  for the case of the teleparallel equivalent of the general relativity,
the horizon of a black hole represents a singular surface for the torsion. 
This latter fact was not noticed in the literature previously. 

Our basic notations and conventions are those of \cite{PR}. In particular,
the signature of the metric is assumed to be $(-,+,+,+)$. Spacetime 
coordinates are labeled by the Latin indices, $i,j,\dots = 0,\dots,3$ (for 
example, $dx^i$), whereas the Greek indices, $\alpha,\beta,\dots = 0,\dots,3$,
label the local frame components (for example, the coframe 1-form  
$\vartheta^\alpha$). Along with the coframe 1-forms $\vartheta^{\alpha}$, we
will widely use the so called $\eta$-basis of the dual coframes. Namely, we
define the Hodge dual such that $\eta:={}^*1$ is the volume 4-form.
Furthermore, denoting by $e_{\alpha}$ the vector frame, we have that
$\eta_{\alpha} := e_{\alpha}\rfloor\eta = {}^*\vartheta_{\alpha}$,
$\eta_{\alpha\beta}:= e_{\beta}\rfloor \eta_{\alpha}={}^*(\vartheta_{\alpha}
\wedge \vartheta_{\beta})$, $\eta_{\alpha \beta\gamma}: = e_{\gamma}\rfloor
\eta_{\alpha\beta}$, $\eta_{\alpha\beta\gamma \delta}: = e_{\delta} \rfloor
\eta_{\alpha\beta\gamma}$. The last expression is thus the totally
antisymmetric Levi-Civita tensor.

\section{Preliminaries: Metric-affine approach}\label{mag}

In this section we recall some basic facts concerning metric-affine geometry
in four dimensions. For a more detailed discussion in arbitrary dimensions,
see \cite{PR}. The metric-affine spacetime is described by the metric
$g_{\alpha\beta}$, the coframe 1-forms $\vartheta^{\alpha}$, and the linear
connection 1-forms $\Gamma_{\beta}{}^{\alpha}$. These are interpreted as
the generalized gauge potentials, while the corresponding field strengths
are the nonmetricity 1-form $Q_{\alpha\beta}=-Dg_{\alpha\beta}$, and the
2-forms of torsion $T^{\alpha}=D\vartheta^{\alpha}$ and curvature 
$R_{\beta}{}^{\alpha}=d\Gamma_{\beta}{}^{\alpha} + \Gamma_{\gamma}{}^{\alpha}
\wedge\Gamma_{\beta}{}^{\gamma}$. 

In our analysis of the teleparallel gravity we will heavily use the
results related to the irreducible decompositions of the fundamental 
geometric and physical objects in MAG. The details about this method
can be found in \cite{PR} and \cite{eff}. Here, we will mainly need
to recall the irreducible parts of the torsion 2-form $T^{\alpha}$ 
which are defined by
\begin{eqnarray}
{}^{(2)}T^{\alpha}&:=&{1\over 3}\vartheta^{\alpha}\wedge T,\quad {\rm with}
\quad T:=e_{\alpha}\rfloor T^{\alpha},\label{T2}\\
{}^{(3)}T^{\alpha}&:=&-\,{1\over 3}{}^*(\vartheta^{\alpha}\wedge P),
\quad {\rm with} \quad
P:={}^*(T^{\alpha}\wedge\vartheta_{\alpha}),\label{T3}\\ {}^{(1)}
T^{\alpha}&:=&T^{\alpha}-{}^{(2)}T^{\alpha} - {}^{(3)}T^{\alpha}.\label{T1}
\end{eqnarray}
In tensor components, the torsion 2-form 
\begin{equation}\label{tor}
T^\alpha = {\frac 1 2}\,T_{ij}{}^\alpha\,dx^i\wedge dx^j 
= {\frac 1 2}\,T_{\mu\nu}{}^\alpha\,\vartheta^\mu\wedge\vartheta^\nu
\end{equation}
gives rise, through the above decomposition, to the vectors of trace 
and axial trace of torsion:
\begin{equation}
e_\alpha\rfloor T = T_{\mu\alpha}{}^\mu,\qquad e_\alpha\rfloor P = 
{\frac 1 2}\,T^{\mu\nu\lambda}\,\eta_{\mu\nu\lambda\alpha}.\label{TP}
\end{equation}

Given the metric $g_{\alpha\beta}$, the Christoffel connection 
$\widetilde{\Gamma}_{\beta}{}^{\alpha}$ is defined as a unique connection 
with vanishing torsion and nonmetricity: $\widetilde{D}\vartheta^{\alpha} 
=0$ and $\widetilde{D}g_{\alpha\beta} =0$. The {\it Riemannian} operators
and  geometrical objects, constructed from the Christoffel connection, will
be  denoted by a tilde. The general affine connection can always be
decomposed  into Riemannian and post-Riemannian parts,
\begin{equation}
\Gamma_{\beta}{}^{\alpha}=\widetilde{\Gamma}_{\beta}{}^{\alpha} 
+ N_{\beta}{}^{\alpha},\label{decom}
\end{equation}
where the {\it distortion} 1-form $N_{\alpha\beta}$ can be 
expressed in terms of torsion and nonmetricity as
\begin{equation}\label{N}
N_{\alpha\beta}=-e_{[\alpha}\rfloor T_{\beta]} + {1\over 2}(e_{\alpha}
\rfloor e_{\beta}\rfloor T_{\gamma})\vartheta^{\gamma} + (e_{[\alpha}
\rfloor Q_{\beta]\gamma})\vartheta^{\gamma} +{1\over 2}Q_{\alpha\beta}.
\end{equation}

The MAG Lagrangian of the gravitational field is constructed from
the forms of curvature, torsion, and nonmetricity. In \cite{eff} the 
method of irreducible decompositions was applied to the most general
quadratic Lagrangian, revealing the possibility of reformulating the
MAG field equations as an effective Einstein equation. Here we will
use the results of \cite{eff} to obtain an analogous reformulation of
teleparallel gravity.

\section{Field equations}\label{eqs}

In the teleparallel theory, we have two geometrical constraints:
\begin{eqnarray}
R_{\beta}{}^{\alpha} &=& d\Gamma_\beta{}^\alpha + \Gamma_\gamma
{}^\alpha\wedge\Gamma_\beta{}^\gamma = 0,\label{Rzero}\\ 
Q_{\alpha\beta} &=& -\,Dg_{\alpha\beta} = - dg_{\alpha\beta} +
\Gamma_\beta{}^\gamma\,g_{\alpha\gamma} + \Gamma_\alpha{}^\gamma
\,g_{\alpha\gamma} = 0.\label{Qzero}
\end{eqnarray}
These equations mean that there is a distant parallelism in spacetime.
The result of a parallel transport of a vector does not depend on the path. 
Moreover, the lengths and angles are not changed during a parallel transport. 

One may wonder, what actually we do gain when we ``embed'' teleparallel
gravity into the MAG theory. After all, the equation (\ref{Rzero}) may be
solved trivially by performing a linear transformation of the frame
and connection
\begin{equation}
e_\alpha\longrightarrow L^\beta{}_\alpha\,e_\beta,\qquad
\Gamma_\alpha{}^\beta\longrightarrow L^\gamma{}_\alpha\Gamma_\gamma{}^\delta
L^{-1\,\beta}{}_\delta + L^{-1\,\beta}{}_\gamma dL^\gamma{}_\alpha.  
\end{equation}
In view of (\ref{Rzero})-(\ref{Qzero}), it is always possible to choose
the matrix $L^\beta{}_\alpha$ of the linear transformation in such a way
that the transformed local metric becomes the standard Minkowski, 
$g_{\alpha\beta} = o_{\alpha\beta} := {\rm diag}(-1, +1, +1, +1)$, whereas 
the transformed local connection becomes trivial: $\Gamma_\alpha{}^\beta
= 0$. Then the (co)frame is left over as the only dynamical variable.
We will call such a choice a {\it Weitzenb\"ock gauge}. The visible
disadvantage of such an approach is the resulting rigid structure of
the tetrad field which cannot be altered at one's wish. To a great 
extent, this is contrary to the very spirit of relativity theory, which
treats all coordinates and frame systems on an equal footing basis. By
embedding teleparallelism into MAG, we remove the unwarranted rigidity of the
tetrad field and return to the far more physically natural situation, where
we can choose a reference and coordinate system at our best convenience. 

Now we can formulate teleparallel gravity as the MAG model with the
constraints (\ref{Rzero})-(\ref{Qzero}). Correspondingly, the Lagrangian 
of the teleparallel theory quite generally can be written as
\begin{equation}
V = {\frac{1}{2\kappa}}\,T^\alpha\wedge{}^*\!\left(\sum_{I=1}^{3}
a_{I}\,^{(I)}T_\alpha\right) + {\frac 12}\,\mu^{\alpha\beta}\wedge
Q_{\alpha\beta} - \nu^\alpha{}_\beta\wedge R_\alpha{}^\beta\,.\label{Ltele}
\end{equation}
Here $a_1, a_2, a_3$ are the three dimensionless coupling constants, and
the gravitational coupling constant is, as usual, $\kappa = 8\pi G/c^3$, 
with $G$ the Newton's gravitational constant. 
Inserting (\ref{T2})-(\ref{tor}) into the first term in (\ref{Ltele}), we find
\begin{equation}\label{TT2}
T^\alpha\wedge{}^*\!\left(\sum_{I=1}^{3}a_{I}\,^{(I)}T_\alpha\right)
= \left(c_1\,T_{\mu\nu}{}^\alpha\,T^{\mu\nu}{}_\alpha + c_2\,
T_{\mu\alpha}{}^\mu\,T^{\nu\alpha}{}_\nu + c_3\,T_{\mu\nu}{}^\alpha
\,T_\alpha{}^{\mu\nu}\right)\eta,
\end{equation}
where the new coefficients are the following combinations of the 
original coupling constants:
\begin{equation}
c_1 = {\frac {2a_1 + a_3} 3},\qquad c_2 = {\frac {2(a_2 - a_1)} 3},
\qquad c_3 = {\frac {2(a_3 - a_1)} 3}.\label{ccc}
\end{equation}

Variation of the teleparallel action with respect to the Lagrange 
multipliers $\mu^{\alpha\beta}$ and $\nu^\alpha{}_\beta$ yields the 
constraints (\ref{Rzero}) and (\ref{Qzero}) of the geometry of distant
parallelism. The Lagrangian (\ref{Ltele}) is invariant under the redefinition
\begin{eqnarray}
\mu^{\alpha\beta} &\longrightarrow&\mu^{\alpha\beta} 
+ D\xi^{\alpha\beta},\label{muxi}\\
\nu^\alpha{}_\beta &\longrightarrow& \nu^\alpha{}_\beta + 
D\chi^\alpha{}_\beta - \xi^\alpha{}_\beta \label{nuchi}
\end{eqnarray}
of the Lagrange multiplier fields. Here, $\xi^{\alpha\beta}=\xi^{\beta\alpha}$
is an arbitrary symmetric 2-form, whereas $\chi^\alpha{}_\beta$ is an
arbitrary 1-form. The proof of the  invariance is based on the three Bianchi
identities of MAG. Indeed, we  have for the additional $\chi$-term arising
from (\ref{nuchi}) in the  Lagrangian:
\begin{equation}
(D\chi^\alpha{}_\beta)\wedge R_\alpha{}^\beta = d\left(\chi^\alpha{}_\beta
\wedge R_\alpha{}^\beta\right) + \chi^\alpha{}_\beta\wedge DR_\alpha{}^\beta.
\end{equation}
Being a total differential, the first term in the right-hand side can be
discarded. The last term vanishes in view of the Bianchi identity
$DR_\alpha{}^\beta\equiv 0$. As for the $\xi$-terms which appear in the
Lagrangian when we make the transformation (\ref{muxi})-(\ref{nuchi}), we
find:
\begin{equation}
{\frac 12}\,(D\xi^{\alpha\beta})\wedge Q_{\alpha\beta} + 
\xi^\alpha{}_\beta\wedge R_\alpha{}^\beta ={\frac 12}\,d\left(
\xi^{\alpha\beta}\wedge Q_{\alpha\beta}\right) - {\frac 12}\,
\xi^{\alpha\beta}\wedge\left(DQ_{\alpha\beta} - 2R_{(\alpha\beta)}\right).
\end{equation}
Again, as a total derivative, the first term can be neglected, whereas the
last term vanishes by virtue of the Bianchi identity $DQ_{\alpha\beta}
\equiv 2R_{(\alpha\beta)}$. As a result, the field equations derived from
(\ref{Ltele}) will determine the Lagrange multipliers only up to the
ambiguity imposed by the symmetry (\ref{muxi})-(\ref{nuchi}). 

The gravitational field equations are derived from the total
Lagrangian $V + L_{\rm mat}$ by independent variations with
respect to the coframe $\vartheta^{\alpha}$ and connection 
$\Gamma_{\beta}{}^{\alpha}$ 1-forms. The corresponding so-called 
{\it first} and {\it second} field equations read 
\begin{eqnarray} 
DH_{\alpha}- E_{\alpha}&=&\Sigma_{\alpha}\,,\label{first}\\ 
DH^{\alpha}{}_{\beta}-E^{\alpha}{}_{\beta}&=&\Delta^{\alpha}{}_{\beta}\,.
\label{second}
\end{eqnarray} 
The covector-valued forms appearing in the left-hand side of 
(\ref{first}) are given by
\begin{eqnarray}
H_{\alpha} &:=& -\,{\partial V \over \partial T^{\alpha}} = -\,{1\over\kappa}
\,{}^*\!\left(\sum_{I=1}^{3}a_{I}{}^{(I)}T_{\alpha}\right),\label{Ha}\\
E_{\alpha} &:=& e_{\alpha}\rfloor V + (e_{\alpha}\rfloor T^{\beta})
\wedge H_{\beta} = {\frac 1 2}\left[(e_{\alpha}\rfloor T^{\beta})
\wedge H_{\beta} - T^\beta\wedge (e_{\alpha}\rfloor H_{\beta})\right].
\end{eqnarray}
In the last expression we have taken into account that $V = -\,{\frac 1 2}
\,T^\alpha\wedge H_\alpha$, in view of (\ref{Ha}) and (\ref{Ltele}).

The matter sources in the right-hand sides of (\ref{first}) and 
(\ref{second}) are the canonical energy-momentum current and the
canonical spin current. For the minimally coupled matter field 
$\Psi^A$ ($p$-form, in general), we have explicitly:
\begin{eqnarray}
\Sigma_\alpha &:=& {\frac {\partial L_{\rm mat}} {\partial\vartheta^\alpha}}
= e_\alpha\rfloor L_{\rm mat} - (e_\alpha\rfloor D\Psi^A)\wedge {\frac 
{\partial L_{\rm mat}} {\partial D\Psi^A}} + (e_\alpha\rfloor\Psi^A)\wedge 
{\frac {\partial L_{\rm mat}} {\partial\Psi^A}},\label{emcan}\\
\Delta^\alpha{}_\beta &:=& {\frac {\partial L_{\rm mat}} 
{\partial\Gamma_\alpha{}^\beta}} = (\ell^\alpha{}_\beta{}^A{}_B\,\Psi^B)
\wedge {\frac {\partial L_{\rm mat}} {\partial D\Psi^A}}.\label{spincan}
\end{eqnarray}
We will assume that the matter field $\Psi^A$ realizes some representation
of the Lorentz group, and then $\ell_{\alpha\beta} = -\,\ell_{\beta\alpha}$
are the corresponding generators of infinitesimal Lorentz transformations. 
Accordingly, the hypermomentum current reduces to the spin current,
\begin{equation}
\Delta_{\alpha\beta} = \tau_{\alpha\beta} = -\,\tau_{\beta\alpha}.
\end{equation}

The tensor-valued forms appearing in the left-hand side of the field 
equation (\ref{second}) are given by
\begin{eqnarray}
M^{\alpha\beta} &:=& -\,2{\partial V \over \partial Q_{\alpha\beta}}
= \mu^{\alpha\beta},\\
H^{\alpha}{}_{\beta} &:=& - {\partial V \over \partial
R_{\alpha}{}^{\beta}} = \nu^\alpha{}_\beta,\\
E^{\alpha}{}_{\beta} &:=& - \vartheta^{\alpha}\wedge H_{\beta} - 
M^{\alpha}{}_{\beta} = - \vartheta^{\alpha}\wedge H_{\beta} - 
\mu^{\alpha}{}_{\beta}.\label{Eab}
\end{eqnarray}
Substituting all this into (\ref{second}), we find the equation
\begin{equation}
\mu^{\alpha}{}_{\beta} + D\nu^{\alpha}{}_{\beta} = - \vartheta^{\alpha}
\wedge H_{\beta} + \Delta^{\alpha}{}_{\beta}.\label{secondMU}
\end{equation}
The left-hand side of this equation is evidently invariant under the
transformations (\ref{muxi})-(\ref{nuchi}). Correspondingly, the field
equation (\ref{secondMU}) offers maximum of the possible: It determines
the gauge invariant piece of the Lagrange multipliers, namely 
$\mu^{\alpha}{}_{\beta} + D\nu^{\alpha}{}_{\beta}$, in terms of the
spin current $\Delta^{\alpha}{}_{\beta}$ and of the translational field 
momentum $H_\alpha$.

It is important to notice that the Lagrange multipliers $\mu^{\alpha\beta}$
and $\nu^\alpha{}_\beta$ decouple from the first field equation
(\ref{first}).  As a result, technically we can simply discard the second
field equation (\ref{second}) because it merely determines the Lagrange
multipliers, whereas the dynamics of the gravitational field is
governed by the equation (\ref{first}).

\section{Effective Einstein equation}\label{ein}

In \cite{eff} it was demonstrated that a certain quadratic metric-affine
Lagrangian has very special properties. This Lagrangian reads
\begin{eqnarray}
V^{(0)}&=& {1 \over 2\kappa}\Bigg\{- R_{\alpha\beta}\wedge
\eta^{\alpha\beta} - {}^{(1)}T^{\alpha}\wedge{}^{*(1)}T_{\alpha}
+2{}^{(2)}T^{\alpha}\wedge{}^{*(2)}T_{\alpha} +
{1\over 2}{}^{(3)}T^{\alpha}\wedge{}^{*(3)}T_{\alpha}\nonumber\\
&&+ {}^{(2)}Q_{\alpha\beta}\wedge\vartheta^{\beta}\wedge
{}^{*(1)}T^{\alpha} - 2{}^{(3)}Q_{\alpha\beta}\wedge
\vartheta^{\beta}\wedge{}^{*(2)}T^{\alpha} -2{}^{(4)}Q_{\alpha\beta}
\wedge\vartheta^{\beta}\wedge{}^{*(2)}T^{\alpha}\nonumber\\
&&+ {1\over 4}{}^{(1)}Q_{\alpha\beta}\wedge{}^{*(1)}Q^{\alpha\beta}
-{1\over 2}{}^{(2)}Q_{\alpha\beta}\wedge{}^{*(2)}Q^{\alpha\beta}
-{1\over 8}{}^{(3)}Q_{\alpha\beta}\wedge{}^{*(3)}Q^{\alpha\beta}
\nonumber\\
&&+{3\over 8}{}^{(4)}Q_{\alpha\beta}\wedge{}^{*(4)}Q^{\alpha\beta}
+ ({}^{(3)}Q_{\alpha\gamma}\wedge\vartheta^{\alpha})\wedge
{}^*({}^{(4)}Q^{\beta\gamma}\wedge\vartheta_{\beta})\Bigg\}.\label{V0}
\end{eqnarray}
The definition of the four irreducible parts of the nonmetricity ${}^{(J)}
Q_{\alpha\beta}$, $J=1,\dots,4$ is given in \cite{PR,eff}; we do not
need them here though because the linear and quadratic nonmetricity terms 
are anyway zero, in view of the teleparallel constraint (\ref{Qzero}).

The gravitational gauge field momenta for the Lagrangian (\ref{V0}) 
are given by
\begin{equation}
H^{(0)}_{\alpha}:=-{\partial V^{(0)}\over \partial T^{\alpha}}\equiv 
-\,{1\over 2\kappa}\,N^{\mu\nu}\wedge\eta_{\alpha\mu\nu},\quad\quad
H^{(0)\alpha}{}_{\beta}:= -{\partial V^{(0)}\over\partial
R_{\alpha}{}^{\beta}} ={1\over 2\kappa}\,\eta^{\alpha}{}_{\beta},
\end{equation}
and the corresponding field equations coincide completely with Einstein's
equations of general relativity:
\begin{equation}
DH^{(0)}_{\alpha}- E^{(0)}_{\alpha}\equiv {1\over 2\kappa}\,
\widetilde{R}^{\mu\nu}\wedge\eta_{\alpha\mu\nu},\quad\quad
DH^{(0)\alpha}{}_{\beta}- E^{(0)\alpha}{}_{\beta}\equiv 0.\label{ident}
\end{equation}

We can use this fact and {\it identically} rewrite the teleparallel 
Lagrangian (\ref{Ltele}) as the sum
\begin{equation}
V = -\,a_1\,V^{(0)} + \widehat{V},\label{split}
\end{equation}
where
\begin{equation} 
\widehat{V} = \frac{1}{2\kappa}\left(\alpha_2\,T^\alpha\wedge{}^*
{}^{(2)}T_\alpha + \alpha_3\,T^\alpha\wedge{}^*{}^{(3)}T_\alpha\right) 
+ {\frac 12}\,\mu^{\alpha\beta}\wedge Q_{\alpha\beta} - \nu^\alpha{}_\beta
\wedge R_\alpha{}^\beta\,,\label{Ltele1}
\end{equation} 
and the new ``shifted'' coupling constants are defined by
\begin{equation}
\alpha_2=a_2 + 2a_1,\qquad \alpha_3=a_3 + {1\over 2}\,a_1.\label{aa}
\end{equation}
Correspondingly, the field momenta can be rewritten in the form
\begin{eqnarray}
&&M^{\alpha\beta} = -\,a_1\,M^{(0)\alpha\beta} + \widehat{M}^{\alpha\beta},
\quad H_{\alpha} = -\,a_1\,H^{(0)}_{\alpha} + \widehat{H}_{\alpha}, 
\quad H^{\alpha}{}_{\beta} = -\,a_1\,H^{(0)\alpha}{}_{\beta} 
+ \widehat{H}^{\alpha}{}_{\beta}, \\ 
&&E_{\alpha} = -\,a_1\,E^{(0)}_{\alpha} + \widehat{E}_{\alpha},\qquad
E^{\alpha}{}_{\beta} = -\,a_1\,E^{(0)\alpha}{}_{\beta} 
+ \widehat{E}^{\alpha}{}_{\beta}.\label{Edec}
\end{eqnarray}
We easily find that $\widehat{M}^{\alpha\beta}:=-2\partial\widehat{V}/\partial 
Q_{\alpha\beta} = M^{\alpha\beta}$, and $\widehat{H}^{\alpha}{}_{\beta}:= 
- \partial\widehat{V}/\partial R_{\alpha}{}^{\beta}= H^\alpha{}_\beta$,
whereas 
\begin{eqnarray}
\widehat{H}_{\alpha} &:=& -\,{\partial\widehat{V}\over\partial T^{\alpha}}
= - \,{1\over\kappa}\,{}^*\left(\alpha_2\,{}^{(2)}T_{\alpha} +
\alpha_3\,{}^{(3)}T_{\alpha}\right)\nonumber\\
&=& -\,{\frac 1 {3\kappa}}\left[\,\alpha_2\,{}^*(\vartheta_\alpha\wedge T)
+ \alpha_3\,\vartheta_\alpha\wedge P\right].\label{hatH}
\end{eqnarray}
Here, we explicitly used the irreducible decompositions (\ref{T2}) and
(\ref{T3}) of the torsion, in terms of the trace 1-form $T$ and the axial 
trace 1-form $P$. Then, with the help of the identities (\ref{ident}), one can
transform the field equations (\ref{first}) and (\ref{second}) of MAG into
\begin{eqnarray} 
{-a_1\over 2}\,\widetilde{R}^{\mu\nu}\wedge\eta_{\alpha\mu\nu} 
&=& \kappa\left(\Sigma_{\alpha} - D\widehat{H}_{\alpha} 
+ \widehat{E}_{\alpha}\right), \label{first1}\\ 
\mu^{\alpha}{}_{\beta} + D\nu^{\alpha}{}_{\beta} &=& - \vartheta^{\alpha}
\wedge\widehat{H}_{\beta} + \Delta^{\alpha}{}_{\beta}.\label{secondMU1}
\end{eqnarray}
Accordingly, we can view the teleparallel field equations as the Einstein
general relativity theory (\ref{first1}) with the effective energy-momentum
current $\Sigma_{\alpha}^{\rm eff} = {\frac 1 {-a_1}}(\Sigma_{\alpha} 
- D\widehat{H}_{\alpha} + \widehat{E}_{\alpha})$. Recall also that
$\widehat{E}_{\alpha} = e_\alpha\rfloor\widehat{V} + (e_\alpha\rfloor
T^\beta)\wedge\widehat{H}_\beta$ and $D\widehat{H}_{\alpha} = \widetilde{D}
\widehat{H}_{\alpha} - N_\alpha{}^\beta\wedge\widehat{H}_\beta$. Now,
substituting (\ref{hatH}) and (\ref{N}), we find explicitly:
\begin{eqnarray}
D\widehat{H}_\alpha - \widehat{E}_\alpha &=& {\frac {\alpha_2}{3\kappa}}
\left[-\,\eta_{\alpha\beta}\wedge\widetilde{D}\,(e^\beta\rfloor T) +
{}^*T_\alpha\wedge T -\,{\frac 12}\,\vartheta_\alpha\wedge 
P\wedge T -{\frac 12}\,e_\alpha\rfloor(T\wedge{}^*T)\right]\nonumber\\
&& +\,{\frac {\alpha_3}{3\kappa}}\left[\vartheta_\alpha\wedge dP - 
2T_\alpha\wedge P + P\wedge e_\alpha\rfloor {}^*P + {\frac 12}\,e_\alpha
\rfloor (P\wedge{}^*P)\right].\label{dH}
\end{eqnarray}

The effective Einstein equation (\ref{first1}) contains a symmetric 
and an antisymmetric parts. It is convenient to consider them 
separately. The antisymmetric piece is extracted by taking the
interior product of $e^\alpha\rfloor$ with (\ref{first1}). Taking into
account (\ref{dH}), the result reads:
\begin{equation}
2\alpha_3\,dP - \alpha_2\left({}^*dT + P\wedge T\right) + e_\alpha\rfloor
\left(\alpha_2{}^*T^\alpha\wedge T - 2\alpha_3\,T^\alpha\wedge P\right)
= 3\kappa\,e^\alpha\rfloor\Sigma_\alpha.\label{dP}
\end{equation}
We can now subtract the antisymmetric part from the equation (\ref{first1}),
which technically means substituting $dP$ from the above equation into
the effective Einstein equation. As a result, we finally obtain
\begin{eqnarray}
{-a_1\over 2}\,\widetilde{R}^{\mu\nu}\wedge\eta_{\alpha\mu\nu} &=&
\kappa\left(\Sigma_{\alpha} - {\frac 1 2}\,\vartheta_\alpha\wedge
e^\beta\rfloor\Sigma_\beta\right) + {\frac {\alpha_2} 3}\,\bigg[
\,\eta_{\alpha\beta}\wedge\widetilde{D}\,(e^\beta\rfloor T) 
- {\frac 1 2}\,\vartheta_\alpha\wedge{}^*dT\nonumber\\
&& -\,{}^*T_\alpha\wedge T + {\frac 1 2}\,\vartheta_\alpha\wedge 
e_\beta\rfloor({}^*T^\beta\wedge T) + {\frac 12}\,e_\alpha\rfloor
(T\wedge{}^*T)\bigg]\nonumber\\
&& +\,{\frac {\alpha_3}3}\!\left[2T_\alpha\wedge\! P - \vartheta_\alpha
\wedge e_\beta\rfloor(T^\beta\wedge P) - P\!\wedge e_\alpha\rfloor {}^*P 
- {\frac 12}\,e_\alpha\rfloor (P\wedge{}^*P)\right].\label{first2}
\end{eqnarray}

Analogously to the separation of the antisymmetric part, we can also extract 
the trace of the effective Einstein field equation. For this purpose, we 
multiply (\ref{first2}) with $\vartheta^\alpha\wedge$ from the left, and we
find:
\begin{equation}
\alpha_2\left(\,d{}^*T + T\wedge{}^*T\right) + {\frac {\alpha_3}3}\,
P\wedge{}^*P = \kappa\,\vartheta^\alpha\wedge\Sigma_\alpha - a_1\,R\,\eta.
\end{equation}
Here, $R:=R_{\alpha\beta}{}^{\beta\alpha}$ is the curvature scalar, and we
used the identities: $\vartheta^\alpha\wedge{}^*T_\alpha = -\,{}^*T$,
$\vartheta^\alpha\wedge\vartheta^\beta\wedge\eta_{\mu\nu} = 2\left(
\delta_\mu^\alpha\delta_\nu^\beta -\delta_\nu^\alpha\delta_\mu^\beta
\right)\eta$. Other useful formulas are: $e_\alpha\rfloor {}^*\psi = {}^*
(\psi\wedge\vartheta_\alpha)$ for any form $\psi$; also $\vartheta^\alpha
\wedge\eta_\beta = \delta_\beta^\alpha\,\eta$, $\vartheta^\alpha\wedge
\eta_{\mu\nu} = \delta^\alpha_\nu\,\eta_\mu - \delta^\alpha_\mu\,\eta_\nu$,
$\vartheta^\alpha\wedge\eta_{\mu\nu\lambda}=\delta^\alpha_\mu
\,\eta_{\nu\lambda} + \delta^\alpha_\nu\,\eta_{\lambda\mu} + 
\delta^\alpha_\lambda\,\eta_{\mu\nu}$.

\section{General relativity limit: a problem with spinning matter?}\label{lim}

When the coupling constants are chosen as
\begin{equation}
a_1 = -\,1,\qquad a_2 = 2,\qquad a_3 = {1\over 2},\label{aGR}
\end{equation}
we find from (\ref{aa}) that all $\alpha_I = 0$. [In terms of the tensor
reformulation (\ref{TT2}), the relations (\ref{ccc}) yield $c_1 = -
\,{\frac 1 2}, c_2 = 2, c_3 = 1$. These are the well-known values of the 
teleparallel equivalent of GR, as used in \cite{jose1,jose2,jose3}, e.g.].
Consequently, $\widehat{V} = 0$, and thus $\widehat{H}_{\alpha} = 0$ 
and $\widehat{E}_{\alpha} =0$. The teleparallel field equations 
(\ref{first1}) reduce to the general relativity theory, except
for the fact that the physical source in the right-hand side of 
(\ref{first1}) is not the ``metrical'', but the  canonical energy-momentum
current $\Sigma_\alpha$. Thus one should be careful when the matter field has
nontrivial spin. 

In order to check the consistency of the teleparallel theory, we first
notice that (\ref{secondMU1}) forces the spin current to satisfy 
\begin{equation}
\tau_{\alpha\beta} = D\nu_{[\alpha\beta]}.
\end{equation}
Consequently, it must be conserved:
\begin{equation}
D\tau_{\alpha\beta} = DD\nu_{[\alpha\beta]} = 0.\label{dtau}
\end{equation}
As the next step, we multiply the Einstein equation (\ref{first1}) from the
left by $\vartheta_\beta$. Then, making an antisymmetrization, we find 
that the antisymmetric piece of the energy-momentum current must vanish: 
\begin{equation}
\vartheta_{[\beta}\wedge\Sigma_{\alpha]} = 0.
\end{equation}
This is a consequence of the symmetry of the Einstein tensor which
is equivalent to the condition $\widetilde{R}^{\mu\nu}\wedge
\vartheta_{[\beta}\wedge\eta_{\alpha]\mu\nu} =0$. The same conclusion
is obtained directly from the equation (\ref{dP}), in which $\alpha_I=0$
leads to the vanishing left-hand side. Recalling now the angular momentum 
conservation law,
\begin{equation}
\vartheta_{[\beta}\wedge\Sigma_{\alpha]} + D\tau_{\beta\alpha} = 0,
\end{equation}
we conclude that the spin current is separately conserved, in full agreement
with (\ref{dtau}).

Summarizing, the teleparallel gravity can consistently couple either to
a spinless matter or to a matter with a conserved spin tensor. Since, for
example, the Dirac spinor field does not have such properties, the
teleparallel description of gravity is not applicable to that case.

\subsection{``New general relativity''}

The ``new general relativity'' \cite{HN,HS,nashed,toma1,toma2} model 
is defined by choosing the coupling constants as follows:
\begin{equation}
a_1 = -\,1,\qquad a_2 = 2,\qquad a_3 \neq {1\over 2}.\label{anew}
\end{equation}
With this choice, the above inconsistency problem is avoided. 

Indeed, now we have $\alpha_3 \neq 0$, and hence (\ref{dP}) reduces to
the meaningful equation for the axial trace of torsion: 
\begin{equation}
dP -  e_\alpha\rfloor\left(T^\alpha\wedge P\right)
= {\frac {3\kappa}{2\alpha_3}}\,e^\alpha\rfloor\Sigma_\alpha.\label{dP1}
\end{equation}
The right-hand side, which represents the antisymmetric part of the
energy-momentum current, can be nontrivial now. 

A different type of inconsistency which arises for this model was first 
noticed by Kopczy\'nski \cite{wojtek} who has shown the existence of an
``extra symmetry'' of the Lagrangian which deforms the coframe (without
touching the connection) in such a way that the axial trace remains
invariant. Such a symmetry makes the theory physically 
non-predictable because torsion is not determined uniquely by the field
equations. Later, Nester \cite{nester} clarified that point by establishing
conditions under which such a hidden symmetry can arise.

\section{Conformal transformations in the teleparallel gravity}\label{co}

Conformal transformation in gravity is usually understood as the scaling
of the line element
\begin{equation}
ds^2 \longrightarrow \Omega^2\,ds^2,\qquad {\rm or\ equivalently,}\qquad
g_{ij} \longrightarrow \Omega^2\,g_{ij}.\label{confmet1}
\end{equation}
The conformal factor $\Omega$ can be an arbitrary function of the spacetime
coordinates. In teleparallel gravity, the above transformation is 
naturally realized as the scaling of the coframe:
\begin{equation}
\vartheta^\alpha\longrightarrow\Omega\,\vartheta^\alpha.\label{confcof}
\end{equation}
In terms of the components $\vartheta^\alpha = \vartheta^\alpha_i\,dx^i$, the
conformal  transformation reads:
\begin{equation}
\vartheta^\alpha_i\longrightarrow\Omega\,\vartheta^\alpha_i,\qquad
{\rm and~accordingly}\qquad g_{ij}=\vartheta^\alpha_i\vartheta^\beta_j
\,g_{\alpha\beta}\longrightarrow\Omega^2\,g_{ij}.\label{confmet2}
\end{equation}
The local metric, which in the case of the orthonormal frames is equal
to the Minkowski metric $g_{\alpha\beta} = o_{\alpha\beta} := {\rm diag}
(-1, +1, +1, +1)$, is not  affected by the scaling. 
The conformal transformation of the coframe (\ref{confcof}) induces the 
scaling of the volume 4-form, $\eta := \vartheta^{\widehat{0}}\wedge
\vartheta^{\widehat{1}}\wedge\vartheta^{\widehat{2}}\wedge
\vartheta^{\widehat{3}}\,\rightarrow\,\Omega^4\eta$. The dual frame 
vectors $e_\alpha$ evidently transform to $\Omega^{-1}e_\alpha$
under (\ref{confcof}).

It is natural to assume that the local linear connection 
$\Gamma_\beta{}^\alpha$ is conformally invariant. This is what we 
encounter in the realization of conformal symmetry in the Poincar\'e
gauge gravity \cite{conf1,conf2,conf3}; see also the discussion
in \cite{ilya}. Then we conclude that the Weitzenb\"ock conditions
(\ref{Rzero}) and (\ref{Qzero}) are not changed by the conformal
transformation. Consequently, (\ref{confcof}) leaves the fundamental
structure of the teleparallel gravity untouched, and it describes a map
between the different teleparallel models. 

Unlike curvature and nonmetricity, the torsion 2-form (which plays the
role of the gauge field strength in the teleparallel theory) transforms 
in a nontrivial way under the conformal scaling. Under the action of 
(\ref{confcof}), we find
\begin{equation}
T^\alpha = D\vartheta^\alpha \longrightarrow \Omega T^\alpha 
+ d\Omega\wedge\vartheta^\alpha.\label{conftor}
\end{equation}
As a result, the torsion trace 1-form changes as $T\,\rightarrow\,\Omega T
- 3\,d\Omega$, and hence only the second part of torsion (\ref{T2}) has 
the nontrivial 
transformation property
\begin{equation}
{}^{(2)}T^{\alpha} \longrightarrow \Omega\,{}^{(2)}T^{\alpha} + 
d\Omega\wedge\vartheta^{\alpha}, \label{confT2}
\end{equation}
whereas the two other irreducible parts of torsion, given by (\ref{T3}) and
(\ref{T1}), transform covariantly:
\begin{equation}
{}^{(1)}T^{\alpha} \longrightarrow \Omega\,{}^{(1)}T^{\alpha},\qquad
{}^{(3)}T^{\alpha} \longrightarrow \Omega\,{}^{(3)}T^{\alpha}.\label{confT13}
\end{equation}
Accordingly, the class of models (\ref{Ltele}), with a vanishing coupling 
constant $a_2 =0$, will display the proper conformal behavior in the sense 
that the teleparallel Lagrangian is rescaled as
\begin{equation}
V \longrightarrow \Omega^2\,V.\label{confV}
\end{equation}
Notice that, since the action is changed, it is therefore {\it not} conformal 
invariant. 

We can use the general Lagrange-Noether machinery \cite{PR} developed 
for MAG to derive the corresponding properties of an arbitrary 
model with a proper conformal behavior. Let us consider a model with
the most general Lagrangian $L = L(\psi, D\psi, g_{\alpha\beta}, 
\vartheta^\alpha, \Gamma_\alpha{}^\beta, T^\alpha)$ for an arbitrary 
matter field $\psi$ interacting with the teleparallel gravitational field. 
Under the action of the above defined (infinitesimal) conformal 
transformation, the Lagrangian changes by 
\begin{equation}
\delta L = \delta\vartheta^\alpha\wedge\Sigma_\alpha + \delta\psi\wedge 
{\frac {\partial L}{\partial \psi}} + d\left[\delta\vartheta^\alpha\wedge
{\frac {\partial L}{\partial T^\alpha}} + \delta\psi\wedge 
{\frac {\partial L}{\partial D\psi}}\right].\label{confDL}
\end{equation}
Here, we took into account the conformal invariance of the metric and
connection, $\delta g_{\alpha\beta}=0$ and $\delta\Gamma_\alpha{}^\beta =0$.
Suppose now that the matter field and the Lagrangian have the proper
conformal behavior in the sense that, under the infinitesimal rescaling
($\Omega = 1 + \omega$), we have
\begin{equation}
\delta\psi = k\,\omega\,\psi,\qquad \delta L = \ell\,\omega\,L,
\end{equation}
with the numbers $k$ and $\ell$ giving the conformal weight of the matter 
field and of the Lagrangian, respectively. Substituting this into 
(\ref{confDL}), we find the two identities:
\begin{eqnarray}
\vartheta^\alpha\wedge{\frac {\partial L}{\partial T^\alpha}} + k\,\psi
\wedge{\frac {\partial L}{\partial D\psi}} = 0,\\
\ell\, L - \vartheta^\alpha\wedge\Sigma_\alpha + \vartheta^\alpha\wedge
D {\frac {\partial L}{\partial T^\alpha}} - T^\alpha\wedge {\frac 
{\partial L}{\partial T^\alpha}} - D\psi\wedge {\frac {\partial L}
{\partial D\psi}} - k\,{\frac {\delta L}{\delta\psi}} = 0.
\end{eqnarray}
These arise, as usual \cite{PR}, by considering the terms proportional
to $\omega$ and $d\omega$ separately. 

Specializing now to the case of the purely gravitational Lagrangian
$L = V$, which does not depend on $\psi$ and has the conformal weight
$\ell =2$ [see Eq.\ (\ref{confV})], we obtain the two identities
\begin{equation}
\vartheta^\alpha\wedge{\frac {\partial V}{\partial T^\alpha}} =0,\qquad
2\,V = \vartheta^\alpha\wedge{\frac {\partial V}{\partial\vartheta^\alpha}} 
+ T^\alpha\wedge{\frac {\partial V}{\partial T^\alpha}}.\label{Lell}
\end{equation}
Turning to the teleparallel model (\ref{Ltele}) under consideration,
we can verify explicitly that these identities are indeed valid for the
case $a_2 =0$ (note that $H_\alpha = -\,\partial V/\partial T^\alpha$).

Every extra invariance of the Lagrangian means that there is a certain 
arbitrariness in the classical solutions of the field equations. In 
particular, the above analysis shows that the solutions of the teleparallel 
models with $a_2 =0$ can only be determined up to an arbitrary scale factor
$\Omega$ of the coframe field. Correspondingly, in order to have a 
predictable teleparallel theory, we will mainly confine ourselves to the 
class of Lagrangians with $a_2\neq 0$.

\section{Spherical symmetry and geometric invariants}\label{ansatz}

Let us now proceed with the analysis of the classical solutions of the
general teleparallel model. As a first step, we naturally turn our
attention to the compact object configurations, and specifically
to the case of spherical symmetry. It is worthwhile to note that
such a study was never performed in full generality for the models
with the three arbitrary coupling constants $a_1, a_2, a_3$. A partial
analysis was done in \cite{HS}.

As usual in the study of exact solutions, we have two complementary
aspects. The first one concerns the convenient choice of the local 
coordinates and of the corresponding {\it ansatz} for the dynamical 
fields. The second aspect is to provide the invariant characterization 
of the resulting geometry. Roughly speaking, the choice of an ansatz 
helps to solve the field equations more easily, whereas the invariant
description provides the correct understanding of the physical 
contents of a solution. 

We look for a spherically symmetric solution with the line element
\begin{equation}\label{ds2}
g = -\,A^2\,dt^2 + B^2\,(dr^2 + r^2\,d\theta^2 + r^2\sin^2\theta\,d\phi^2).
\end{equation}
The two functions $A = A(r)$ and $B = B(r)$ depend on the radial variable
$r$. It is, however, not so trivial to come up with the {\it ansatz for 
a tetrad}. A convenient choice reads:
\begin{equation}
\vartheta^{\widehat{0}} = A\,dt,\quad \vartheta^{\widehat{1}} 
= B\,dr,\quad \vartheta^{\widehat{2}} = Br\,d\theta,\quad 
\vartheta^{\widehat{3}} = Br\sin\theta\,d\phi.\label{cof}
\end{equation}
In addition, for the (non-Riemannian) connection we choose
\begin{equation}
\Gamma_2{}^1 = -\,\Gamma_1{}^2 = -\,d\theta,\quad 
\Gamma_3{}^1 = -\,\Gamma_1{}^3 = -\,\sin\theta\,d\phi,\quad 
\Gamma_3{}^2 = -\,\Gamma_2{}^3 = -\,\cos\theta\,d\phi.\label{gam}
\end{equation}
It is easy to see that (\ref{gam}) is the pure gauge configuration, and 
the curvature is indeed vanishing, $R_{\beta}{}^{\alpha}=d\Gamma_{\beta}
{}^{\alpha} + \Gamma_{\gamma}{}^{\alpha}\wedge\Gamma_{\beta}{}^{\gamma}=0$.
One can certainly perform a linear transformation which yields the 
Weitzenb\"ock gauge, but the resulting tetrad {\it ansatz} becomes somewhat
obscure, see \cite{jose4}, e.g.. This point demonstrates the convenience
of the metric-affine approach, which offers a greater flexibility in 
the choice of the {\it ansatz} for a solution. 

Hereafter, derivatives with respect to the radial coordinate will be denoted
by a prime. Although the curvature is zero, the torsion of the configuration
(\ref{cof})-(\ref{gam}) is nontrivial and reads:
\begin{equation}
T^{\widehat{0}} = -\,{\frac {A'}{AB}}\,\vartheta^{\widehat{0}}\wedge
\vartheta^{\widehat{1}},\quad T^{\widehat{2}} = {\frac {B'}{B^2}}
\,\vartheta^{\widehat{1}}\wedge\vartheta^{\widehat{2}},\quad 
T^{\widehat{3}} = {\frac {B'}{B^2}}\,\vartheta^{\widehat{1}}\wedge
\vartheta^{\widehat{3}}.\label{torsphere}
\end{equation}
Correspondingly, the irreducible pieces of torsion are given by
\begin{equation}
{}^{(1)}T^\alpha = -\,{\frac {(A'B - B'A)}{3AB^2}}\,\left(
2\,\delta^\alpha_{\widehat{0}}\,\vartheta^{\widehat{0}}\wedge
\vartheta^{\widehat{1}} + \delta^\alpha_{\widehat{2}} 
\,\vartheta^{\widehat{1}}\wedge\vartheta^{\widehat{2}} +
\delta^\alpha_{\widehat{3}}\,\vartheta^{\widehat{1}}\wedge
\vartheta^{\widehat{3}}\right),
\end{equation}
and
\begin{equation}
{}^{(2)}T^\alpha = {\frac {(A'B + 2B'A)}{3AB^2}}\,\left(
-\,\delta^\alpha_{\widehat{0}}\,\vartheta^{\widehat{0}}\wedge
\vartheta^{\widehat{1}} + \delta^\alpha_{\widehat{2}} 
\,\vartheta^{\widehat{1}}\wedge\vartheta^{\widehat{2}} +
\delta^\alpha_{\widehat{3}}\,\vartheta^{\widehat{1}}\wedge
\vartheta^{\widehat{3}}\right).
\end{equation}
The axial torsion vanishes identically: ${}^{(3)}T^\alpha =0$.

The above {\it ansatz} is clearly a coordinate- and frame-dependent 
statement. In order to have a correct understanding of the resulting
solution, we need to construct invariants of the curvature and
torsion. The total (Riemann-Cartan) curvature is identically zero in 
the teleparallel gravity. However, the {\it Riemannian} curvature of 
the Christoffel connection for the metric (\ref{ds2}) is nontrivial. 
In particular, computation of the Weyl 2-form yields:
\begin{eqnarray}
\widetilde{W}^{\widehat{0}\widehat{1}} &=&\ {\frac W3}
\,\vartheta^{\widehat{0}}\wedge\vartheta^{\widehat{1}},\qquad 
\widetilde{W}^{\widehat{0}\widehat{2}} = -\,{\frac W6}
\,\vartheta^{\widehat{0}}\wedge \vartheta^{\widehat{2}},\qquad 
\widetilde{W}^{\widehat{0}\widehat{3}} = -\,{\frac W6}
\,\vartheta^{\widehat{0}}\wedge \vartheta^{\widehat{3}},\\
\widetilde{W}^{\widehat{1}\widehat{2}} &=& -\,{\frac W6}
\,\vartheta^{\widehat{1}}\wedge\vartheta^{\widehat{2}},\qquad 
\widetilde{W}^{\widehat{3}\widehat{1}} = -\,{\frac W6}
\,\vartheta^{\widehat{3}}\wedge \vartheta^{\widehat{1}},\qquad 
\widetilde{W}^{\widehat{2}\widehat{3}} = {\frac W3}
\,\vartheta^{\widehat{2}}\wedge \vartheta^{\widehat{3}},\label{Weyl}
\end{eqnarray}
with
\begin{equation}
W = {\frac r {AB}}\left[{\frac 1 r}
\left({\frac A B}\right)^\prime\right]^\prime.\label{Winv}
\end{equation}
The components of this 2-form represent the Weyl tensor, 
$\widetilde{W}^{\alpha\beta} = {\frac 1 2}\,C_{\mu\nu}{}^{\alpha\beta}
\,\vartheta^\mu\wedge\vartheta^\nu$. The Weyl quadratic invariant thus reads 
\begin{equation}
\widetilde{W}_{\alpha\beta}\wedge{}^*\widetilde{W}^{\alpha\beta} 
= {\frac {W^2}3}\,\eta,
\end{equation}
and consequently we can consistently use (\ref{Winv}) for the 
description of the resulting geometry. 

Besides the Riemannian Weyl tensor, the spacetime geometry is naturally
characterized by the quadratic invariants of the torsion. For the
spherically symmetric configurations (\ref{torsphere}), we have explicitly:
\begin{equation}
T_\alpha\wedge{}^\ast T^\alpha = {\frac 1 {B^2}}\left[\left({\frac {A'}A}
\right)^2 - 2\left({\frac {B'}B}\right)^2\right]\eta.\label{tortor}
\end{equation}
These two invariants --- the Riemannian curvature and the quadratic torsion 
--- provide the sufficient tools for understanding the contents of the
classical solutions.

\section{Coupled gravitational and electromagnetic fields}\label{emf}

As we discovered above, the material sources with spin may lead 
to certain inconsistencies in the framework of teleparallelism.
Correspondingly, in order to be on the safe side, we will limit
ourselves to the case of the spinless matter when the teleparallel
gravity appears to be totally applicable. Of all the possible 
spinless matter, the Maxwell field clearly represents a very 
interesting and physically important case. 

Accordingly, in our study of the spherically symmetric solutions, we will 
investigate the case when matter is represented by the electromagnetic 
field. The spin current of the electromagnetic field is trivial, whereas 
its energy-momentum reads:
\begin{equation}
\Sigma_\alpha = {\frac \lambda 2}\,\left[(e_\alpha\rfloor F)\wedge 
{}^\ast F - (e_\alpha\rfloor {}^\ast F)\wedge F\right].\label{Sigma}
\end{equation}
Here, the electromagnetic vacuum constant (the ``vacuum impedance'')
\begin{equation}
\lambda = \sqrt{\frac {\varepsilon_0}{\mu_0}}, \label{kaplam}
\end{equation}
is defined in terms of the electric $\varepsilon_0$ and magnetic $\mu_0$
constants of the vacuum. 

Using the standard spherically symmetric {\it ansatz} for the electromagnetic 
potential 1-form, 
\begin{equation}
{\cal A} = {\frac f A}\,\vartheta^{\widehat{0}} = f\,dt,\label{Apot}
\end{equation}
with $f=f(r)$, we have for the field strength
\begin{equation}
F =d{\cal A} = -\,{\frac {f'}{AB}}\,\vartheta^{\widehat{0}}\wedge
\vartheta^{\widehat{1}},\qquad {}^\ast F = {\frac {f'}{AB}}
\,\vartheta^{\widehat{2}}\wedge\vartheta^{\widehat{3}}.\label{FdualF}
\end{equation}
Then, we derive the Maxwell equation:
\begin{equation}
d\,{}^\ast\!F = {\frac 1 {r^2B^3}}\,{\frac {d}{dr}}\left\{{\frac 
{r^2B} A}\,f'\right\}\,\vartheta^{\widehat{1}}\wedge
\vartheta^{\widehat{2}}\wedge\vartheta^{\widehat{3}} = 0.\label{Maxw}
\end{equation}

On the other hand, a direct computation yields for the left-hand side of the
teleparallel gravitational field equations:
\begin{eqnarray}
DH_{\widehat{0}} - E_{\widehat{0}} &=& {\frac {(U_0 + U_1 - 2U_2)}
{6\kappa B^2}}\,\vartheta^{\widehat{1}}\wedge\vartheta^{\widehat{2}}
\wedge\vartheta^{\widehat{3}},\label{lhs0}\\
DH_{\widehat{1}} - E_{\widehat{1}} &=& {\frac {U_1}{6\kappa B^2}}
\,\vartheta^{\widehat{0}}\wedge\vartheta^{\widehat{2}}\wedge
\vartheta^{\widehat{3}},\label{lhs1}\\
DH_{\widehat{2}} - E_{\widehat{2}} &=& {\frac {(U_2 - U_1)}{6\kappa B^2}}
\,\vartheta^{\widehat{0}}\wedge\vartheta^{\widehat{3}}\wedge
\vartheta^{\widehat{1}},\label{lhs2}\\
DH_{\widehat{3}} - E_{\widehat{3}} &=& {\frac {(U_2 - U_1)}{6\kappa B^2}}
\,\vartheta^{\widehat{0}}\wedge\vartheta^{\widehat{1}}\wedge
\vartheta^{\widehat{2}}.\label{lhs3}
\end{eqnarray}
Here, we have denoted
\begin{eqnarray}
U_0 &:=& {\frac 2 {ABr^2}}\,{\frac d {dr}}\left\{ ABr^2
\left[ (4a_1 - a_2)\,{\frac {A'} A} - 2(2a_1 + a_2)\,{\frac {B'} B}
\right]\right\},\label{U0}\\
U_1 &:=& {\frac 4 r}\left( (a_1 - a_2)\,{\frac {A'} A} 
- (a_1 + 2a_2)\,{\frac {B'} B} \right) \nonumber\\
&& - \,(2a_1 + a_2)\left({\frac {A'} A}\right)^2
- 2(a_1 + 2a_2)\left({\frac {B'} B}\right)^2 + 4(a_1 - a_2)
\,{\frac {A'} A}{\frac {B'} B}\,,\label{U1}\\
U_2 &:=& {\frac 2 {ABr^3}}\,{\frac d {dr}}\left\{ ABr^3
\left[ (a_1 - a_2)\,{\frac {A'} A} - (a_1 + 2a_2)\,{\frac {B'} B}
\right]\right\}.\label{U2}
\end{eqnarray}

Substituting (\ref{FdualF}) into (\ref{Sigma}), we get explicitly
the components of the electromagnetic energy-momentum current 3-form:
\begin{eqnarray}
\Sigma_{\widehat{0}} &=& -\,{\frac \lambda 2}\left({\frac 
{f'}{AB}}\right)^2\,\vartheta^{\widehat{1}}\wedge
\vartheta^{\widehat{2}}\wedge\vartheta^{\widehat{3}},\label{Sig0}\\
\Sigma_{\widehat{1}} &=& \quad {\frac \lambda 2}\left({\frac 
{f'}{AB}}\right)^2\,\vartheta^{\widehat{0}}\wedge
\vartheta^{\widehat{2}}\wedge\vartheta^{\widehat{3}},\label{Sig1}\\
\Sigma_{\widehat{2}} &=& -\,{\frac \lambda 2}\left({\frac 
{f'}{AB}}\right)^2\,\vartheta^{\widehat{0}}\wedge
\vartheta^{\widehat{3}}\wedge\vartheta^{\widehat{1}},\label{Sig2}\\
\Sigma_{\widehat{3}} &=& -\,{\frac \lambda 2}\left({\frac 
{f'}{AB}}\right)^2\,\vartheta^{\widehat{0}}\wedge
\vartheta^{\widehat{1}}\wedge\vartheta^{\widehat{2}}.\label{Sig3}
\end{eqnarray}
In the presence of a nontrivial electromagnetic field, we need, in 
addition to the above geometric invariants, an invariant description 
of the matter source configuration. As it is well known, there are 
two invariants of the Maxwell field. For the spherical {\it ansatz}
(\ref{FdualF}), one of the invariants is trivial, $F\wedge F \equiv 0$,
whereas the other reads
\begin{equation}
F\wedge {}^*F = - \left({\frac {f'}{AB}}\right)^2\,\eta.\label{FF0}
\end{equation}

\section{Analyzing the field equations}\label{an}

The Maxwell equation (\ref{Maxw}) can be straightforwardly integrated.
This yields
\begin{equation}
f' = {\frac {qA}{r^2B}},\label{Af1}
\end{equation}
with $q$ an integration constant. Its value is determined by the total 
electric charge $Q$ of the source which is calculated as usual from
the integral over the 2-sphere around the source:
\begin{equation}
Q = \int\limits_{S_2}\,\lambda\,{}^\ast F =\int\limits_{S_2}\,\lambda
\,{\frac {f'}{AB}}\,\vartheta^{\widehat{2}}\wedge\vartheta^{\widehat{3}}
= \int\limits_{S_2}\,\lambda q\,\sin\theta d\theta\wedge d\phi = 
4\pi\,\lambda q.\label{Qq}
\end{equation}
Inserting (\ref{Af1}) into (\ref{FF0}), we find for the Maxwell invariant
\begin{equation}
F\wedge {}^*F = -\,{\frac {q^2}{r^4B^4}}\,\eta.\label{FF1}
\end{equation}

From (\ref{lhs0})-(\ref{lhs3}) and (\ref{Sig0})-(\ref{Sig3}) we find,
after making use of (\ref{Af1}), and of some simple rearrangements:
\begin{equation}
U_0 = -\,{\frac {6\kappa\lambda q^2}{r^4B^2}},\qquad
U_1 = {\frac {3\kappa\lambda q^2}{r^4B^2}},\qquad U_2 = 0.\label{U012}
\end{equation}
Using (\ref{U0}) in (\ref{U012}), we obtain the equation
\begin{equation}
{\frac d {dr}}\left\{ ABr^2\left[ (4a_1 - a_2)\,{\frac {A'} A} - 2(2a_1 
+ a_2)\,{\frac {B'} B}\right]\right\} = {\frac {-\,3\kappa\lambda q^2A}
{r^2B}} = -\,3\kappa\lambda q\,f',
\end{equation}
where we have used (\ref{Af1}) in the last step. Consequently, the first 
integral is straightforwardly obtained:
\begin{equation}
(4a_1 - a_2)\,{\frac {A'} A} - 2(2a_1 + a_2)\,{\frac {B'} B} =
{\frac {k_1 - 3\kappa\lambda q\,f}{r^2AB}}, \label{k1}
\end{equation}
with $k_1$ an integration constant. Analogously, from (\ref{U2}) 
and (\ref{U012}) we find another first integral:
\begin{equation}
(a_1 - a_2)\,{\frac {A'} A} - (a_1 + 2a_2)\,{\frac {B'} B} =
{\frac {k_2}{r^3AB}},\label{k2}
\end{equation}
with $k_2$ a second integration constant.

Let us introduce a new variable by
\begin{equation}
\varphi := k_1 - 3\kappa\lambda qf.\label{fnew}
\end{equation}
Differentiating this and using (\ref{Af1}), we find the equation
\begin{equation}
\varphi' = -\,{\frac {3\kappa\lambda q^2\, A}{r^2\,B}}.\label{Af2}
\end{equation}
Suppose $a_1a_2\neq 0$ (the special case $a_1a_2 =0$ will be considered
separately). Then, combining (\ref{k1}) and (\ref{k2}), we finally get the 
system of first order equations
\begin{eqnarray}
{\frac {A'} A} &=& {\frac 1 {9a_1a_2r^2AB}}\left[(a_1 + 2a_2)\,\varphi
- 2(2a_1 + a_2){\frac {k_2} r}\right],\label{Adot}\\
{\frac {B'} B} &=& {\frac 1 {9a_1a_2r^2AB}}\left[(a_1 - a_2)\,\varphi
- (4a_1 - a_2){\frac {k_2} r}\right].\label{Bdot}
\end{eqnarray}
Together with (\ref{Af2}), these equations comprise a system of three first
order ordinary differential equations for the three unknown functions $A, B,
f$. As an immediate consequence, the sum of (\ref{Adot}) and (\ref{Bdot})
yields 
\begin{equation}
(AB)' = {\frac 1 {9a_1a_2r^2}}\left[(2a_1 + a_2)\,\varphi
- (8a_1 + a_2){\frac {k_2} r}\right].\label{ABdot}
\end{equation}

For the sake of completeness, we should recall that there is one more 
equation, which is derived by using (\ref{U1}) in (\ref{U012}):
\begin{eqnarray}
{\frac {3\kappa\lambda q^2}{r^4B^2}} &=& {\frac 4 r}\left( (a_1 - a_2)
\,{\frac {A'} A} - (a_1 + 2a_2)\,{\frac {B'} B} \right) \nonumber\\
&-& \,(2a_1 + a_2)\left({\frac {A'} A}\right)^2
- 2(a_1 + 2a_2)\left({\frac {B'} B}\right)^2 + 4(a_1 - a_2)
\,{\frac {A'} A}{\frac {B'} B}\,.\label{U1a}
\end{eqnarray}
However, this last equation is satisfied automatically for the solutions
(\ref{Af2})-(\ref{Bdot}) of the system. The easiest way to see this is 
to substitute (\ref{Adot})-(\ref{Bdot}) into (\ref{U1a}). We then find
\begin{equation}
(a_1 + 2a_2)\,\varphi^2 - 4(2a_1 + a_2)\,{\frac {k_2}r}\,\varphi
+ 2(8a_1 + a_2)\,{\frac {k_2^2}{r^2}} - 9a_1a_2\,(4k_2\,AB -
3\kappa\lambda q^2\,A^2) = 0.\label{add}
\end{equation}
Now, if we differentiate this equation, the result is identically zero
by virtue of the equations (\ref{Af2}), (\ref{ABdot}) and (\ref{Adot}). 
However, since not only the equation itself vanishes for the solutions,
but also its derivative, we still have to keep this equation. Ultimately 
it will turn out that (\ref{add}) imposes certain relation between 
the various integration constants.

\section{Conformally flat solutions}\label{flat}

Let us study a very special case, when the metric functions are
proportional. Namely, 
\begin{equation}
A = a_0\,B,\label{BaA}
\end{equation}
with a constant coefficient $a_0$. Then, the metric (\ref{ds2}) 
describes the conformally flat spacetime geometry. The pair of 
equations (\ref{k1}) and (\ref{k2}) reduce to the system
\begin{eqnarray}
3a_2{\frac {B'} B} &=& -\,{\frac {\varphi}{a_0r^2B^2}},\label{k1a0}\\
3a_2{\frac {B'} B} &=& -\,{\frac {k_2}{a_0r^3B^2}}.\label{k2a0}
\end{eqnarray}
Notice that the Lagrangian with the vanishing coupling constant $a_2 =0$ 
belongs to the class of the conformally covariant teleparallel gravity
models  [see Sec.~\ref{co}]. As a result, the overall factor of the tetrad,
and hence of the metric, is undetermined. This is manifested in our 
spherically symmetric case as well: If we put $a_2=0$ in Eqs.\
(\ref{k1a0})-(\ref{k2a0}), we find a vanishing electromagnetic field 
$\varphi =0$, whereas the conformal metric factor $B$ remains completely 
arbitrary. 

We will assume that the teleparallel model is {\it not} conformally
covariant. Then $a_2\neq 0$, and the system (\ref{k1a0})-(\ref{k2a0})
yields the explicit electromagnetic function
\begin{equation}
\varphi = {\frac {k_2} r}.
\end{equation}
Substituting this, together with (\ref{BaA}), into (\ref{Af2}), we get
\begin{equation}
a_0 = {\frac {k_2}{3\kappa\lambda q^2}}.\label{a0}
\end{equation}
Furthermore, the integration of (\ref{k2a0}) yields
\begin{equation}
B^2 = {\frac {\kappa\lambda q^2}{a_2r^2}} + k_3.
\end{equation}
The new integration constant is fixed to be equal to zero $k_3 = 0$, which
follows from Eq.\ (\ref{U1a}) after all the above is substituted.
Thus, we obtain finally the conformally flat solution 
\begin{equation}
A^2 = {\frac {k_2^2}{9\kappa\lambda q^2a_2\,r^2}},\qquad
B^2 = {\frac {\kappa\lambda q^2}{a_2\,r^2}}.\label{ABa0}
\end{equation}

In order to understand this spacetime geometry, we have to compute
the corresponding Riemannian curvature (recall that the {\it total} 
Riemann-Cartan curvature is identically zero). A direct calculation yields
\begin{equation}
\widetilde{R}^{\widehat{0}\widehat{1}} = {\frac {a_2}{\kappa\lambda q^2}}
\,\vartheta^{\widehat{0}}\wedge\vartheta^{\widehat{1}},\qquad 
\widetilde{R}^{\widehat{2}\widehat{3}} = -\,{\frac {a_2}{\kappa\lambda q^2}}
\,\vartheta^{\widehat{2}}\wedge\vartheta^{\widehat{3}},\label{curBR}
\end{equation}
with all other components of the curvature 2-form trivial. As we see, the
curvature is everywhere regular, and the Riemann tensor has the 
double-duality property,
\begin{equation}
\widetilde{R}_{\alpha\beta}{}^{\gamma\delta} = {\frac 1 4}
\,\eta^{\gamma\delta\rho\sigma}\,\eta_{\alpha\beta\mu\nu}
\,\widetilde{R}_{\rho\sigma}{}^{\mu\nu}.
\end{equation}
The Riemannian Weyl tensor vanishes identically, as it is clearly seen from 
Eq.\ (\ref{Weyl}). This is consistent with the fact that the metric is
conformally  flat. Geometrically, the resulting spacetime is the direct
product of the  two 2-dimensional spaces of constant curvature, a hyperbolic
space and a sphere. A solution of that type was originally described by
Bertotti and  Robinson \cite{bertrob1,bertrob2} in the framework of
general relativity theory.

In the teleparallel gravity, we have torsion as the basic field variable.
It is straightforward to see that in the generalized Bertotti-Robinson
solution (\ref{ABa0}), torsion is also constant with the quadratic
invariant (\ref{tortor}) given by
\begin{equation}\label{torBR}
T_\alpha\wedge{}^\ast T^\alpha = -\,{\frac {a_2}{\kappa\lambda q^2}}\,\eta.
\end{equation}
Note that the constant magnitude of the torsion is again described by
the same combination of the coupling constants which determine the
constant Riemannian curvature (\ref{curBR}) of that solution.

The electromagnetic invariant (\ref{FF1}), when using (\ref{ABa0}), 
demonstrates that the source is likewise represented by the ``constant'' 
electromagnetic field configuration:
\begin{equation}\label{FFconf}
F\wedge {}^*F = -\left({\frac {a_2}{\kappa\lambda q}}\right)^2\,\eta.
\end{equation}

\subsection{Vanishing coupling constant $\hbox{$a_1 =0$}$}

Before we proceed to consider the general case, we have to analyze the
special case when $a_1a_2 =0$, see Sec.~\ref{an}. When $a_2 =0$, we 
have the conformally covariant theory, and consequently, the conformal 
factor of the metric remains undetermined. We will not consider such 
models which lack physical predictability. 

When, however, $a_1 =0$, the pair of equations (\ref{k1}) and  (\ref{k2})
reduce to the system
\begin{eqnarray}
{\frac {A'} A} + 2{\frac {B'} B} &=& -\,{\frac {\varphi}{a_2r^2AB}},\\
{\frac {A'} A} + 2{\frac {B'} B} &=& -\,{\frac {k_2}{a_2r^3AB}}.\label{k2a1}
\end{eqnarray}
This system yields the explicit electromagnetic function
\begin{equation}
\varphi = {\frac {k_2} r}.
\end{equation}
Substituting this into (\ref{Af2}), we find the proportionality of the
metric functions:
\begin{equation}
B = {\frac {3\kappa\lambda q^2}{k_2}}\,A.\label{ABa1}
\end{equation}
This brings us then back to the results of the previous subsection.

\section{Uncharged solutions}\label{un}

{}From now on, we will confine our attention to the generic teleparallel
models with $a_1a_2\neq 0$. Let us first obtain the configurations with zero
charge, $q = 0$.  Then, from (\ref{fnew}) we have $\varphi = k_1$, and the
equation (\ref{ABdot}) can be immediately integrated:
\begin{equation}
AB = {\frac 1 {9a_1a_2r}}\left[ -\,(2a_1 + a_2)\,k_1
+ (8a_1 + a_2){\frac {k_2} {2r}}\right] + k_3.\label{ABsol}
\end{equation}
When we substitute this into the equation (\ref{add}), we find the 
value of the new integration constant explicitly: 
\begin{equation}
k_3 = {\frac {(a_1 + 2a_2)\,k_1^2}{36a_1a_2\,k_2}}.\label{k3}
\end{equation}

\subsection{Special case: $8a_1 + a_2 = 0$}\label{excase}

Suppose the coupling constants satisfy
\begin{equation}
8a_1 + a_2 = 0.
\end{equation}
Then, (\ref{ABsol}) and (\ref{k3}) yield
\begin{equation}
AB = {\frac {k_1}{12a_1}}\left({\frac {5k_1}{8k_2}} - {\frac 1 r}\right).
\end{equation}
Using this in Eq.\ (\ref{Adot}), we derive the ordinary differential
equation:
\begin{equation}
{\frac {A'}{A}} = {\frac {2k_2}{k_1\,r^2}}\left(1 - {\frac 
{\frac {5k_1}{8k_2}} {{\frac 1 r} - {\frac {5k_1}{8k_2}}}}\right).
\end{equation}
The integration is straightforward, and the final solution for the
metric function can be written in the form:
\begin{eqnarray}
A^2 &=& A_0^2\,e^{-\,{\frac {5r_0}{2r}}}\left({\frac {r_0}{r}} 
- 1\right)^{5/2},\label{A0sing}\\
B^2 &=& B_0^2\,e^{\frac {5r_0}{2r}}\left({\frac {r_0}{r}} 
- 1\right)^{-\,1/2}.\label{B0sing}
\end{eqnarray}
Here, $A_0, B_0$ and $r_0$ are arbitrary constants. For the sake of
completeness,   let us give their relation to the original (coupling and
integration)  constants:
\[
r_0 = 8k_2/5k_1, \quad A_0B_0 = 5k_1^2/96k_2a_1.
\] 

This spacetime is {\it not a black hole}. The curvature tensor is singular 
at $r=0$ and at $r=r_0$, so although the metric component $g_{00} = -A^2$ 
vanishes at the finite radius $r=r_0$, this is not a horizon. Indeed, 
substituting $A$ and $B$ into (\ref{Weyl}), we get the Riemannian Weyl
curvature:
\begin{equation}
W = -\,{\frac {r_0\,\left(48r^3  - 136r^2r_0 + 110rr_0^2  - 25r_0^3  
\right)}{4r^6\,B_0^2\,e^{\frac {5r_0}{2r}}(r_0/r - 1)^{\frac 3 2}}}.
\end{equation}
It is easy to see that it diverges at $r=r_0$. Analogously, for the torsion
invariant  (\ref{tortor}), we find:
\begin{equation}\label{torsp}
T_\alpha\wedge{}^\ast T^\alpha = {\frac {r_0^2\,\left(172r^4 - 564r^3r_0 
+ 687r^2r_0^2 - 370rr_0^3  + 75r_0^4\right)}{16r^8\,B_0^2\,e^{\frac {5r_0}
{2r}}(r_0/r - 1)^{\frac 7 2}}}\,\eta.
\end{equation}
Thus, from both Riemannian and teleparallel viewpoints, the resulting 
geometry is singular at $r=0$ and at $r=r_0$.

\subsection{General case: $8a_1 + a_2 \neq 0$}

When $8a_1 + a_2 \neq 0$, we can use (\ref{k3}) to rewrite (\ref{ABsol}) as 
\begin{equation}\label{ABsol2}
AB = {\frac {k_1^2(8a_1 + a_2)}{18a_1a_2k_2}}\left({\frac {k_2}{k_1\,r}} - 
{\frac \alpha 2}\right)\left({\frac {k_2}{k_1\,r}} - {\frac \beta 2}\right),
\end{equation}
where we have introduced the constant parameters
\begin{eqnarray}
\alpha&=&{\frac {2(2a_1 + a_2) + \sqrt{-\,18a_1a_2}}{8a_1 + a_2}},\label{al}\\
\beta &=&{\frac {2(2a_1 + a_2) - \sqrt{-\,18a_1a_2}}{8a_1 + a_2}}.\label{be}
\end{eqnarray}
In particular, we can easily see that
\begin{equation}
{\frac {\alpha + \beta} 2} = {\frac {2(2a_1 + a_2)}{8a_1 + a_2}},\qquad
{\frac {\alpha\beta} 4} = {\frac {a_1 + 2a_2}{2(8a_1 + a_2)}}.
\end{equation}
{}From (\ref{al})-(\ref{be}) we conclude that the product $a_1a_2$ must
be negative. Substituting (\ref{ABsol2}) into (\ref{Adot}), we find the 
equation
\begin{equation}
{\frac {A'} A} = -\,{\frac {k_2}{k_1\,r^2}}\left[{\frac \alpha 
{{\frac {k_2}{k_1\,r}} - {\frac \alpha 2}}} + {\frac \beta 
{{\frac {k_2}{k_1\,r}} - {\frac \beta 2}}}\right].
\end{equation}
Integration yields
\begin{equation}
A = A_0\left({\frac {k_2}{k_1\,r}} - {\frac \alpha 2}\right)^\alpha
\left({\frac {k_2}{k_1\,r}} - {\frac \beta 2}\right)^\beta.\label{A0sol}
\end{equation}
Using then (\ref{ABsol2}), we find
\begin{equation}
B = B_0\left({\frac {k_2}{k_1\,r}} - {\frac \alpha 2}\right)^{1- \alpha}
\left({\frac {k_2}{k_1\,r}} - {\frac \beta 2}\right)^{1 -\beta}.\label{B0sol}
\end{equation}
The integration constants are originally related by $A_0B_0 = k_1^2(8a_1 + 
a_2)/18a_1a_2k_1$. However, taking into account the possibility of scaling
both the time and the radial coordinate by two arbitrary factors, the
constants $A_0$ and $B_0$ can have any real value. In order to simplify the
notation, it will be convenient to introduce $m = k_2/k_1$. 

Performing a differentiation in (\ref{Weyl}), we find the nontrivial
Riemannian Weyl curvature:
\begin{equation}
W = {\frac {m\,w}{r^6\,B_0^2}}\left({\frac {m}{r}} 
- {\frac \alpha 2}\right)^{2\alpha - 4}\left({\frac {m}{r}} 
- {\frac \beta 2}\right)^{2\beta - 4},\label{WT}
\end{equation}
where we denoted the polynomial
\begin{eqnarray}
w &:=& {\frac 1 8}\,\Big\{32m^3(\alpha + \beta)(\alpha + \beta - 1) 
+ 4m^2r\left[(\alpha + \beta)(3 - 2(\alpha + \beta) - 16\alpha\beta)
+ 12\alpha\beta\right]\nonumber\\
&&\quad +\,2mr^2\left[ 8\alpha\beta(2\alpha\beta  + \alpha + \beta -1)
- (\alpha + \beta)^2\right] + 3r^3\alpha\beta(\alpha + \beta 
- 4\alpha\beta)\Big\}\nonumber\\
&=& {\frac {-\,3\,a_2}{(8a_1 + a_2)^2}}\,\Big[a_1\,(r - 4m)(3r^2 - 8mr + 8m^2) 
+ 2a_2\,(r - m)^2(3r - 8m)\Big].\label{w}
\end{eqnarray}
Analogously, for the quadratic torsion invariant (\ref{tortor}), we find
\begin{equation}
T_\alpha\wedge{}^\ast T^\alpha = {\frac {m^2\,{\cal T}}{r^6\,B_0^2}}
\left({\frac {m}{r}} - {\frac \alpha 2}\right)^{2\alpha - 4}\left(
{\frac {m}{r}} - {\frac \beta 2}\right)^{2\beta - 4}\,\eta,\label{TT}
\end{equation}
where we have another polynomial defined as
\begin{eqnarray}
{\cal T} &:=& m^2\left[3(\alpha + \beta)^2 + 8(1 - \alpha - \beta)\right]
+ 2mr\left[(\alpha + \beta)^2 - (\alpha + \beta)(2 + 3\alpha\beta) 
+ 4\alpha\beta\right]\nonumber\\ &&\qquad\qquad +\,r^2\left[(\alpha + 
\beta)^2/2 + \alpha\beta(3\alpha\beta - 2(\alpha + \beta))\right]\nonumber\\
&=& {\frac {12}{(8a_1 + a_2)^2}}\,\Big[\,a_1^2\,(r - 4m)^2 + 2a_2^2
\,(r - m)^2\Big].\label{Tinv}
\end{eqnarray}

\section{Coupling constants, invariants, and singularities}\label{values}

The values of the constants $\alpha$ and $\beta$ are crucial for 
understanding the spacetime geometry of the solutions obtained 
above, since they determine the behavior of the metric functions 
(\ref{A0sol}) and (\ref{B0sol}). An straightforward analysis of 
(\ref{al})-(\ref{be}) shows that we can simplify those formulas 
to the following equivalent ones:
\begin{eqnarray}
\alpha&=&{\frac {2(a_1+\sqrt{-2a_1a_2})}{4a_1+\sqrt{-2a_1a_2}}},\label{al1}\\
\beta&=&{\frac {2(a_1-\sqrt{-2a_1a_2})}{4a_1-\sqrt{-2a_1a_2}}}.\label{be1} 
\end{eqnarray}
Moreover, one can eliminate $\sqrt{-2a_1a_2}$ from (\ref{al1})-(\ref{be1}), 
and express one constant directly in terms of the other:
\begin{equation}
\alpha = {\frac {4 - 5\beta}{5 - 4\beta}},\qquad {\rm or~ equivalently},
\qquad \beta = {\frac {4 - 5\alpha}{5 - 4\alpha}}.\label{albe}
\end{equation}
{}From these simple derivations, we can immediately establish a number of 
important consequences. Since $a_1a_2 < 0$ (in order to have real solutions),
we have to analyze the two cases for the original coupling constants: 
(a) $a_1> 0, a_2 < 0$, and (b) $a_1< 0, a_2 > 0$.

For $a_1> 0, a_2 < 0$, we see from (\ref{al1}) that $\alpha$ 
is {\it positive}. Moreover,
\begin{equation}
{\frac 1 2} < \alpha < 2\qquad {\rm for\ all}\qquad a_1, a_2.
\end{equation}
At the same time, equation (\ref{be1}) shows that
\begin{eqnarray}
0 <\beta < {\frac 1 2} &&\qquad {\rm for}\qquad a_1 > -\,2a_2,\\
\beta > 2 &&\qquad {\rm for}\qquad a_1 < -\,a_2/8,\\
\beta < 0 &&\qquad {\rm for}\qquad -\,a_2/8 < a_1 < -\,2a_2.
\end{eqnarray}
For $a_1 = -2a_2$, we see that $\beta =0$, whereas when $a_1 = 
-a_2/8$, we return to the exceptional case considered in Sec.~\ref{excase}
(then, $\beta$ formally diverges). 

For $a_1< 0, a_2 > 0$, we analogously find from (\ref{be1}) 
that now $\beta$ is {\it positive},
\begin{equation}
{\frac 1 2} < \beta < 2\qquad {\rm for\ all}\qquad a_1, a_2,
\end{equation}
whereas (\ref{al1}) yields
\begin{eqnarray}
0 <\alpha < {\frac 12} &&\qquad {\rm for}\qquad a_1 < -\,2a_2,\\
\alpha > 2 &&\qquad {\rm for}\qquad a_1 > -\,a_2/8,\\
\alpha < 0 &&\qquad {\rm for}\qquad -\,2a_2 < a_1 < -\,a_2/8.
\end{eqnarray}
For $a_1 = -2a_2$, we see that $\alpha =0$, whereas when $a_1 = 
-a_2/8$, we again obtain the exceptional case considered in Sec.~\ref{excase}
(then $\alpha$ becomes infinite). 

It is important to find the zeros of the metric coefficient $g_{00} = 
-\,A^2$, since these values determine the position of a possible horizon. 
For the solutions under consideration, $A$ always has one or two zeros: 
(a) at $r = 2m/\alpha$ when $\alpha > 0$, and/or (b) at $r=2m/\beta$ if 
$\beta >0$. The function (\ref{A0sol}) would not have zeros only in the 
case when {\it both} $\alpha \leq 0$ and $\beta\leq 0$, but this never 
happens according to the above analysis.

In order to decide whether a zero of $g_{00}$ corresponds to a horizon,
we have to study the behavior of the curvature and torsion invariants
at that value of the radial coordinate. As it is clear from (\ref{WT}), 
the Riemannian curvature 
diverges at the zeros of $A$, unless the polynomial (\ref{w}) also vanishes 
there. Hence, it is important to find the value of the polynomial (\ref{w}) 
at the zeros of the function $A$. A direct substitution yields:
\begin{equation}
w(2m/\alpha) = 2m^3\,(\alpha - \beta)^2{\frac {2\alpha^2 -3\alpha + 1}
{\alpha^2}},\quad w(2m/\beta) = 2m^3\,(\alpha - \beta)^2{\frac 
{2\beta^2 -3\beta + 1}{\beta^2}}.\label{Wab}
\end{equation}
Since $\alpha$ can never be equal to $\beta$ (for any nonzero coupling
constants $a_1, a_2$), we conclude from (\ref{Wab}) that the polynomial 
$w$ can have common zeros with $A$ only when either $\alpha$ or $\beta$ 
is equal to 1 or 1/2. Equations (\ref{al1})-(\ref{be1}) show that this 
is possible only when $2a_1 + a_2 =0$ (then either $\alpha$ or $\beta$ 
is equal 1), or when one of the two constants $a_1$ or $a_2$ vanishes. 

Analogously, for the torsion quadratic invariant we find:
\begin{equation}
{\cal T}(2m/\alpha) = m^2\,(\alpha - \beta)^2{\frac {3\alpha^2 -4\alpha + 2}
{\alpha^2}},\quad {\cal T}(2m/\beta) = m^2\,(\alpha - \beta)^2{\frac 
{3\beta^2 - 4\beta + 2}{\beta^2}}.\label{Tab}
\end{equation}
As one can easily see, these quantities are nonvanishing for any choice 
of the constants $\alpha, \beta$. Correspondingly, the torsion invariant
is always singular at the zeros of $g_{00}$ in all teleparallel gravity
models.

\subsection{General relativity limit: Schwarzschild black hole}

As we have mentioned in Sec.~\ref{lim}, the specific choice (\ref{aGR}) 
of the coupling constants gives rise to a teleparallel model
which is called the teleparallel equivalent of GR. Accordingly,  
when $a_1 = -1$ and  $a_2 = 2$, we find from (\ref{al})-(\ref{be}) that
\begin{equation}
\alpha = -\,1,\qquad \beta = 1.\label{abgr}
\end{equation}
Consequently, the metric functions reduce to
\begin{equation}
A = A_0\left({\frac m r} - {\frac 12}\right)\left({\frac m r} + {\frac 12}
\right)^{-1},\qquad B = B_0\left({\frac m r} + {\frac 12}\right)^2.
\end{equation}
Furthermore, the polynomial (\ref{w}) now reads
\begin{equation}
w = -\,{\frac {3r}2}\,(2m - r)^2 = -\,6r^3\left({\frac m r} 
- {\frac 12}\right)^2,\label{w1}
\end{equation}
and consequently, the Riemannian Weyl curvature is described by
\begin{equation}
W = -\,{\frac {6m\,r^3}{B_0^2\,(m + r/2)^6}}.\label{Wschw}
\end{equation}
The most important observation is that the Riemannian curvature is thus
regular at the zero $r = 2m$ of the metric function $A(r)$, which means
that we have a horizon here. The resulting geometry then describes the 
well known Schwarzschild black hole with the horizon at $r =2m$. 
The singularity at $r = -\,2m$ is the usual point-mass 
source singularity at the origin. 

Since we are dealing with teleparallel gravity, it is necessary also
to analyze the behavior of torsion. A direct substitution of 
(\ref{abgr}) into (\ref{TT})-(\ref{Tinv}) yields:
\begin{equation}
T_\alpha\wedge{}^\ast T^\alpha = {\frac {m^2\,(3r^2 - 8mr + 8m^2)\,r^2}
{B_0^2\,(m + r/2)^6\,(m - r/2)^2}}\,\eta.\label{Tschw}
\end{equation}
As we see, the torsion invariant diverges not only at the origin $r = -\,2m$,
but also at the Schwarzschild horizon $r =2m$. Unlike the ``regularizing''
effect of the polynomial (\ref{w1}) in the curvature invariant, the 
analogous polynomial in the numerator of (\ref{Tschw}) does not help to 
remove the singularity of the torsion invariant. 

We emphasize this fact as the main difference between the standard
general relativistic description of the Schwarzschild black hole and its
teleparallel counterpart. The horizon is a regular surface from the viewpoint
of the Riemannian geometry, but it is singular from the viewpoint of
teleparallel gravity.

\subsection{No black holes in teleparallel gravity?}

Returning to the case of the general teleparallel Lagrangian, we may ask if 
a (class of) model exists with a specific choice of the coupling constants
for which the above solutions describe a black hole.  As we see from
(\ref{WT}), the curvature is singular at both zeros of the  metric function
$A$, i.e., at $r = 2m/\alpha$ {\it and} at $r = 2m/\beta$, when
\begin{equation}
\alpha < 2 \qquad {\rm and}\qquad \beta < 2. \label{ab1}
\end{equation}
Correspondingly, recalling the results of the beginning of the current 
section, the teleparallel Lagrangians with $|a_1| > |a_2|/8$ do not have 
spherically symmetric solutions describing black holes.

There is, in principle, a possibility that the configurations with
\begin{equation}
\alpha > 2 \qquad {\rm or}\qquad \beta > 2 \label{ab2}
\end{equation}
may turn out to be black holes. Note that both $\alpha$ and $\beta$ cannot 
be simultaneously greater than 2, as was shown above in this section.
When these conditions are fulfilled, the curvature becomes regular 
either at $r = 2m/\alpha$ or at $r = 2m/\beta$.
However, the problem is that the corresponding surface ($r = 2m/\alpha$ 
or $r = 2m/\beta$) would have an infinite area. This is different
from what we usually expect from a horizon surface. 

Consequently, we come to the conclusion that {\it for all possible values 
of the coupling constants $a_1$ and $a_2$ --- except for the special case 
$a_1 = -1$ and $a_2 = 2$} --- the uncharged spherically symmetric solutions 
(\ref{A0sing})-(\ref{B0sing}), as well as (\ref{A0sol})-(\ref{B0sol})
of the general teleparallel model (\ref{Ltele}), do not describe black 
hole configurations. In this sense, the so-called teleparallel equivalent of
GR is distinguished among all other teleparallel models as the only
theory which admits black holes.

\section{Charged solutions: general relativity limit}\label{ch}

Let us now return to the charged solutions determined by the system
(\ref{Af2})-(\ref{Bdot}). In view of the above result, one can expect that
the most physically interesting case, within the class of the general
teleparallel theories (\ref{Ltele}), corresponds to the teleparallel
equivalent of GR. Accordingly, we will confine our attention now to 
the case $a_1 = -\,1, a_2 = 2$, or equivalently to (\ref{abgr}).
Then, the equations (\ref{Bdot})-(\ref{ABdot}) are reduced to 
\begin{eqnarray}
{\frac {A'} A} &=& -\,{\frac \varphi {6r^2AB}},\label{AdotGR}\\
{\frac {B'} B} &=& {\frac 1 {6r^2AB}}\left(\varphi - {\frac {2k_2}r}\right),\\
(AB)' &=& -\,{\frac {k_2} {3r^3}}\,.\label{ABdotGR}
\end{eqnarray}
The last equation is easily integrated, yielding
\begin{equation}
AB = {\frac 1 6}\left({\frac {k_2}{r^2}} + k_3\right),\label{ABgr}
\end{equation}
where $k_3$ is a new integration constant.
Combining (\ref{Af2}) and (\ref{AdotGR}), we can eliminate $B$ to find
\begin{equation}
A'A = {\frac {\varphi'\varphi}{18\kappa\lambda q^2}}.
\end{equation}
This immediately yields the first integral
\begin{equation}
A^2 = {\frac {\varphi^2 + k_4}{18\kappa\lambda q^2}}. \label{A2phi}
\end{equation}
The new integration constant $k_4$ can have any sign, as well as be 
equal zero. Each case will be studied separately below. 
Substituting (\ref{A2phi}) and (\ref{ABgr}) into the additional equation
(\ref{add}), we find the following relation between the integration
constants:
\begin{equation}
4k_2k_3 = k_4.\label{kkk}
\end{equation}
The final step is to find the function $\varphi(r)$ explicitly. 
Substituting (\ref{A2phi}) and (\ref{ABgr}) into (\ref{Af2}), and taking into 
account the condition (\ref{kkk}), we obtain the differential equation
\begin{equation}
\varphi' = 
-\,{\frac {k_4 + \varphi^2}{k_2[1 + k_4(r/2k_2)^2]}},\label{Af4}
\end{equation}
The form of the solution depends crucially on the value of $k_4$.

\subsection{Negative constant $k_4$}

For negative integration constant, $k_4 = -\,|k_4|$, the solution
of (\ref{Af4}) reads:
\begin{equation}
\varphi(r) = \sqrt{|k_4|}\,{\frac {[k_5\,(r + r_0)^2 + (r - r_0)^2]}
{[k_5\,(r + r_0)^2 - (r - r_0)^2]}}, \label{phigr}
\end{equation}
where $k_5$ is a new integration constant, and we denoted 
\begin{equation}
r_0 := {\frac {2k_2}{\sqrt{|k_4|}}}.\label{r0}
\end{equation}
Substituting (\ref{phigr}) into (\ref{A2phi}) and (\ref{ABgr}), we find 
the metric functions:
\begin{eqnarray}
A^2 &=& {\frac {2k_5|k_4|}{9\kappa\lambda q^2}}\left[
{\frac {r^2 - r_0^2}{k_5\,(r + r_0)^2 - (r - r_0)^2}}\right]^2,\label{Agr}\\
B^2 &=& {\frac {\kappa\lambda q^2}{32k_5r_0^2\,r^4}}\,\left[
k_5\,(r + r_0)^2 - (r - r_0)^2\right]^2. \label{Bgr}
\end{eqnarray}
In order to have a correct signature of the metric (\ref{ds2}),
we must assume that $k_5> 0$.

As we see, the metric coefficient $g_{00} = -\,A^2$ has zeros at 
$r = \pm r_0$. These values can qualify for the positions of a horizon,
and in order to clarify this we have to study the behavior of the 
geometric and electromagnetic invariants. The quadratic torsion
invariant (\ref{tortor}) is as follows: 
\begin{equation}
T_\alpha\wedge{}^\ast T^\alpha = {\frac {128\,k_5\,r_0^4\,r^2\,\check{\cal T}}
{\kappa\lambda q^2\,(r^2 - r_0^2)^2\,[k_5\,(r + r_0)^2 - (r - r_0)^2]^4}}
\,\eta,\label{TTn}
\end{equation}
where the polynomial reads
\begin{eqnarray}
\check{\cal T} &=& k_5^2\,(3r^2 - 4rr_0 + 2r_0^2)(r + r_0)^4 +
2k_5\,(3r^2 - 2r_0^2)(r^2 - r_0^2)^2\nonumber\\
&&\qquad +\,(3r^2 + 4rr_0 + 2r_0^2)(r - r_0)^4.\label{Tn}
\end{eqnarray}
On the other hand, for the Maxwell invariant (\ref{FF1}) we find
\begin{equation}
F\wedge{}^*F = -\,{\frac {1024\,k_5^2\,r_0^4\,r^4}{(\kappa\lambda q)^2
\,[k_5\,(r + r_0)^2 - (r - r_0)^2]^4}}\,\eta.\label{FFn}
\end{equation}
As far as the Riemannian curvature is concerned, everything becomes more
transparent if we make the coordinate transformation defined by
\begin{eqnarray}
\rho &=& \sqrt{\frac {\kappa\lambda q^2}{8k_5}}\left[{\frac {k_5\,(r + r_0)^2 
- (r - r_0)^2}{2r_0r}}\right],\\
\overline{t} &=& -\,{\frac 1 {3c(k_5 - 1)}}\,\sqrt{\frac {2k_5|k_4|}
{\kappa\lambda q^2}}\,t.
\end{eqnarray}
Then, the line element transforms into the standard Reissner-Nordstr\"om
form: 
\begin{equation}\label{RNmet1}
ds^2  = -\,h^2c^2 d\overline{t}^2 + \frac{1}{h^2}\,d \rho^2
+ \rho^2\left(d\theta^2+\sin^2\theta \,d\phi^2\right),
\end{equation}
with
\begin{equation}\label{func1}
h^2 = 1 - \frac{2Gm}{c^2\rho} +\frac{GQ^2}{4\pi\varepsilon_0c^4\rho^2}.
\end{equation}
The mass parameter is here introduced by
\begin{equation}
{\frac{Gm}{c^2}} = (k_5 + 1)\sqrt{\frac {\kappa\lambda q^2}{8k_5}},
\end{equation}
whereas the charge $Q$ of the source is given, as usual, by (\ref{Qq}).

It is easy to see that the values $r=\pm r_0$ correspond to the two 
zeros of the metric function $h^2(\rho)$. As a result, we conclude that
these values give the position of the horizon because both the Maxwell
invariant (\ref{FFn}) and the Riemannian curvature (as it is well
known) are regular there. However, it is remarkable that the torsion
(\ref{TTn}) is again singular at those surfaces, just like in the
uncharged solutions considered earlier.

\subsection{Positive constant $k_4$}

For a positive integration constant $k_4 = |k_4|$, the integration 
of (\ref{Af4}) is also straightforward, yielding
\begin{equation}
\varphi(r) = \sqrt{|k_4|}\,{\frac {[k_5\,(r^2 - r_0^2) + 2rr_0]}
{[r^2 - r_0^2 - 2k_5rr_0]}}.\label{phigr2}
\end{equation}
Here, $k_5$ is a new integration constant (which is different from the 
constant $k_5$ introduced in the previous subsection), and we use the 
same $r_0$ as defined in (\ref{r0}). We then easily find the metric 
functions, using (\ref{A2phi}) and (\ref{ABgr}):
\begin{eqnarray}
A^2 &=& {\frac {(1 + k_5^2)|k_4|}{18\kappa\lambda q^2}}\left[
{\frac {r^2 + r_0^2}{r^2 - r_0^2 - 2k_5rr_0}}\right]^2,\label{Agr2}\\
B^2 &=& {\frac {\kappa\lambda q^2}{8(1 + k_5^2)r_0^2\,r^4}}
\,\left[r^2 - r_0^2 - 2k_5rr_0\right]^2. \label{Bgr2}
\end{eqnarray}
Now, we may notice that the coordinate transformation
\begin{eqnarray}
\rho &=& \sqrt{\frac {\kappa\lambda q^2}{2(1+k_5^2)}}\left[
{\frac {r^2 - r_0^2 - 2k_5rr_0}{2r_0r}}\right],\\
\overline{t} &=& {\frac 1 {3c}}\,\sqrt{\frac {(1 + k_5^2)|k_4|}
{2\kappa\lambda q^2}}\,t,
\end{eqnarray}
again brings the line element to the Reissner-Nordstr\"om form 
(\ref{RNmet1}), where this time the mass is introduced via
\begin{equation}
{\frac{Gm}{c^2}} =  -\,k_5\sqrt{\frac {\kappa\lambda q^2}{2(1 + k_5^2)}}.
\end{equation}
This shows that $k_5$ should be negative in this case. 

We will not give the curvature, torsion, and Maxwell invariants explicitly,
but their qualitative behavior is the same as in the previous case: The
torsion invariant is again singular on a Riemannian horizon of the 
corresponding Reissner-Nordstr\"om black hole.

\subsection{Vanishing constant $k_4$}

For the sake of completeness, it remains to consider the case of vanishing
integration constant $k_4 =0$. Then, as follows from (\ref{kkk}), one should
also have  either $k_2 =0$ or $k_3 =0$.

\subsubsection{Case: $k_4=0$, $k_2 =0$}

In this case, the equation (\ref{Af4}) is easily solved to give
\begin{equation}
\varphi = {\frac {k_3\,r}{k_5\,r - 1}},\label{phi02}
\end{equation}
with an arbitrary integration constant $k_5$. Correspondingly, (\ref{A2phi})
and (\ref{ABgr}) yield:
\begin{eqnarray}
A^2 &=& {\frac {k_3^2\,r^2}{18\kappa\lambda q^2(k_5\,r - 1)^2}},\\
B^2 &=& {\frac {\kappa\lambda q^2(k_5\,r -1)^2}{2r^2}}. 
\end{eqnarray}
The coordinate transformation 
\begin{eqnarray}
\rho &=& -\,\sqrt{\frac {\kappa\lambda q^2} 2}\,(k_5\,r - 1),\\
\overline{t} &=& {\frac {k_3\,t}{3ck_5\sqrt{2\kappa\lambda q^2}}},
\end{eqnarray}
brings the metric to the extremal Reissner-Nordstr\"om line element
\begin{equation}\label{RNdeg}
ds^2 = -\left(1 - {\frac {\sqrt{\frac {\kappa\lambda q^2} 2}}\rho}\right)^2
d\overline{t}^2 + \left(1 - {\frac {\sqrt{\frac {\kappa\lambda q^2} 2}}\rho}
\right)^{-2}d\rho^2 + \rho^2\left(d\theta^2+\sin^2\theta \,d\phi^2\right)\,.
\end{equation}

\subsubsection{Case: $k_4=0$, $k_3 =0$}

Analogously, the equation (\ref{Af4}) is integrated,
\begin{equation}
\varphi = {\frac {k_2}{r + k_5}},\label{phi03}
\end{equation}
with an integration constant $k_5$. Then, (\ref{A2phi})
and (\ref{ABgr}) yield the metric functions:
\begin{eqnarray}
A^2 &=& {\frac {k_2^2}{18\kappa\lambda q^2(r + k_5)^2}},\\
B^2 &=& {\frac {\kappa\lambda q^2(r + k_5)^2}{2r^4}}. 
\end{eqnarray}
The coordinate transformation 
\begin{eqnarray}
\rho &=& \sqrt{\frac {\kappa\lambda q^2} 2}\left(1 + {\frac {k_5}r}\right),\\
\overline{t} &=& {\frac {k_2\,t}{3ck_5\sqrt{2\kappa\lambda q^2}}},
\end{eqnarray}
again yields the extremal Reissner-Nordstr\"om metric (\ref{RNdeg}).

In both cases, the torsion diverges at $\rho = \sqrt{\kappa\lambda q^2/2}$,
whereas this surface appears to be regular from the Riemannian point of
view.

\section{Discussion and conclusions}

In this paper we have studied the general teleparallel gravity model 
within the framework of the MAG theory. A similar analysis of ``embedding''
teleparallelism into the Poincar\'e gauge theory was performed in 
\cite{wojtek} within the framework of the Lagrangian formalism, and also 
by using the Hamiltonian methods in \cite{ham1,ham2}.
Generalizing the previous work, we consider the full 3-parameter 
teleparallel Lagrangian without {\it a priori} restricting the coupling 
constants $a_1, a_2, a_3$. The main motivation for this was to 
determine the place and significance of the so-called teleparallel
GR-equivalent model which is specialized by the values (\ref{aGR}). 
It is well known that, for obvious reasons, the GR-equivalent 
teleparallel theory is satisfactorily supported by observations.

Our study reveals a new qualitative feature which distinguishes
the teleparallel GR-equivalent among other models: The spherically
symmetric solutions (charged and uncharged) describe black hole 
configurations only for the special choice (\ref{aGR}) of the coupling
constants. We have thus demonstrated that a generic teleparallel model  does
not admit black holes. There exists, though, a family of completely regular
solutions which appears to be a direct generalization of the
Bertotti-Robinson solution. 

Another new result obtained concerns the behavior of the curvature and
torsion invariants in the general teleparallel model. We find that the 
quadratic torsion invariant displays a much worse singularity structure 
than one could expect from the analysis of the Riemannian curvature
invariants. In particular, even in the teleparallel GR-equivalent model, the
black hole solutions have torsion singularities on a horizon surface which
is, however, regular from the point of view of the curvature. This 
striking result raises a question about the geometrical and physical
meaning of the torsion singularities in teleparallel gravity. 

Finally, it seems worthwhile to note that, although the teleparallel 
GR-equivalent model has a number of nice features which distinguishes
it among the general teleparallel theories, it still has a consistency 
problem of coupling of matter with spin to the teleparallel gravitational
field. As a matter of fact, one can argue that there is no such a 
problem because, being a gauge theory of the group of translations,
teleparallelism is thus, by definition, related only to the energy-momentum
current. And indeed, teleparallelism turns out to be completely
consistent for the case of spinless matter, which is characterized solely by
the energy-momentum. From this point of view, teleparallelism 
appears to be not applicable to matter sources with spin, and
our analysis has clearly demonstrated that point. 

\begin{acknowledgments}
The work of YNO was supported by FAPESP. JGP thanks FAPESP and CNPq for
partial financial support.
\end{acknowledgments}

\end{document}